\documentclass[runningheads]{llncs}
\usepackage{booktabs} 
\usepackage{graphicx}
 \usepackage{xcolor}
\usepackage{listings}
\usepackage{footnote}
\usepackage{algorithm}
\usepackage{amssymb}
\usepackage{stmaryrd}
\usepackage{amsmath}

\usepackage{xspace}
\usepackage{paralist}
\usepackage{multirow}
\usepackage{adjustbox} 
\usepackage[inline]{enumitem} 
\usepackage{makecell}
\RequirePackage{tikz}
 \usetikzlibrary{arrows,automata,shapes,calc,through,decorations.pathmorphing,decorations.fractals,chains,shapes.multipart}

 \definecolor{javared}{rgb}{0.6,0,0} 
 \definecolor{javagreen}{rgb}{0.25,0.5,0.35} 
 \definecolor{javapurple}{rgb}{0.5,0,0.35} 
 \definecolor{javadocblue}{rgb}{0.25,0.35,0.75} 

\usepackage{subfigure}
\usepackage{arydshln} 

\usepackage[pdftex]{hyperref}
 \hypersetup{
 pdftitle={},
 pdfauthor={},
 pdfkeywords={},
 colorlinks=true,
 citecolor={purple},
 linkcolor={blue},
 bookmarksopen=true,
 bookmarksnumbered=true,
 }
\pdfoutput=1

\newcommand{\num}[1]{\relax\ifmmode \mathbb #1\else $\mathbb #1$\fi}
\newcommand{\nnnum}[1]{\relax\ifmmode 
  {\mathbb #1}_{\geq 0} \else ${\mathbb #1}_{\geq 0}$
  \fi}
\newcommand{\npnum}[1]{\relax\ifmmode 
  {\mathbb #1}_{\leq 0} \else ${\mathbb #1}_{\leq 0}$
  \fi}
\newcommand{\pnum}[1]{\relax\ifmmode 
  {\mathbb #1}_{> 0} \else ${\mathbb #1}_{> 0}$
  \fi}
\newcommand{\nnum}[1]{\relax\ifmmode 
  {\mathbb #1}_{< 0} \else ${\mathbb #1}_{< 0}$
  \fi}
\newcommand{\plnum}[1]{\relax\ifmmode 
  {\mathbb #1}_{+} \else ${\mathbb #1}_{+}$
  \fi}
\newcommand{\nenum}[1]{\relax\ifmmode 
  {\mathbb #1}_{-} \else ${\mathbb #1}_{-}$
  \fi}

\newcommand{\extb}[1]{\relax\ifmmode {\sf ExtBeh}_{#1} \else ${\sf ExtBeh}_{#1}$\fi} 
\newcommand{\tdists}[1]{\relax\ifmmode {\sf Tdists}_{#1} \else ${\sf Tdists}_{#1}$\fi} 

\newcommand{\exec}[1]{\relax\ifmmode {\sf Execs}_{#1} \else ${\sf Exec}_{#1}$\fi} 
\newcommand{\execf}[1]{\relax\ifmmode {\sf Execs}^*_{#1} \else ${\sf Exec}^*_{#1}$\fi} 
\newcommand{\execi}[1]{\relax\ifmmode {\sf Execs}^\omega_{#1} \else ${\sf Exec}^\omega_{#1}$\fi} 

\newcommand{\ctrace}[1]{\relax\ifmmode {\sf Ctraces}_{#1} \else ${\sf Ctraces}_{#1}$\fi} 

\newcommand{\trace}[1]{\relax\ifmmode {\sf Traces}_{#1} \else ${\sf Traces}_{#1}$\fi} 
\newcommand{\tracef}[1]{\relax\ifmmode {\sf Traces}^*_{#1} \else ${\sf Traces}^*_{#1}$\fi} 
\newcommand{\tracei}[1]{\relax\ifmmode {\sf Traces}^\omega_{#1} \else ${\sf Traces}^\omega_{#1}$\fi} 

\newcommand{\frag}[1]{\relax\ifmmode {\sf Frags}_{#1} \else ${\sf Frags}_{#1}$\fi} 
\newcommand{\fragf}[1]{\relax\ifmmode {\sf Frags}^*_{#1} \else ${\sf Frags}^*_{#1}$\fi} 
\newcommand{\fragi}[1]{\relax\ifmmode {\sf Frags}^\omega_{#1} \else ${\sf Frags}^\omega_{#1}$\fi} 

\newcommand{\reach}[1]{\relax\ifmmode {\sf Reach}_{#1} \else ${\sf Reach}_{#1}$\fi}

\def\H{{\mathcal H}} 
\def\P{{\mathcal P}} 
\def\T{{\cal T}} 
\def\X{{\mathcal X}} 

\newcommand{\col}[1]{\relax\ifmmode \mathscr #1\else $\mathscr #1$\fi}

\definecolor{HIOAcolor}{rgb}{0.776,0.22,0.07}

\newcommand{\SC}[2]{\relax\ifmmode {\tt Scount}(#1,#2) \else ${\tt Scount}(#1,#2)$\fi} 
\newcommand{\SCM}[2]{\relax\ifmmode {\tt Smin}(#1,#2) \else ${\tt Smin}(#1,#2)$\fi} 
\newcommand{\Aut}[1]{\relax\ifmmode {\tt Aut}(#1) \else ${\tt Aut}(#1)$\fi}

\newcommand{\deq}{\mathrel{\stackrel{\scriptscriptstyle\Delta}{=}}}

\newcommand{\seclabel}[1]{\label{sec:#1}}
\newcommand{\secref}[1]{Section~\ref{sec:#1}}

\newcommand{\figlabel}[1]{\label{fig:#1}}
\newcommand{\figref}[1]{Figure~\ref{fig:#1}}

\newcommand{\tabref}[1]{Table~\ref{tab:#1}}
\newcommand{\applabel}[1]{\label{app:#1}}
\newcommand{\appref}[1]{Appendix~\ref{app:#1}}

\renewcommand{\eqref}[1]{Equation~\ref{eq:#1}}

\newcommand{\formlabel}[1]{\label{form:#1}}

\newcommand{\remove}[1]{}
\newcommand{\salg}[1]{\relax\ifmmode {\mathcal F}_{#1}\else ${\mathcal F}_{#1}$\fi} 
\newcommand{\msp}[1]{\relax\ifmmode (#1, \salg{#1}) \else $(#1, \salg{#1})$\fi} 
\newcommand{\msprod}[2]{\relax\ifmmode ( #1 \times #2, \salg{#1} \otimes \salg{#2}) \else $(#1 \times #2, \salg{#1} \otimes \salg{#2})$\fi} 
\newcommand{\dist}[1]{\relax\ifmmode {\mathcal P}\msp{#1}
  \else ${\mathcal P}\msp{#1}$\fi} 
\newcommand{\subdist}[1]{\relax\ifmmode {\mathcal S}{\mathcal P}\msp{#1} 
  \else ${\mathcal S}{\mathcal P}\msp{#1}$\fi} 
\newcommand{\disc}[1]{\relax\ifmmode {\sf Disc}(#1)
  \else ${\sf Disc}(#1)$\fi} 

\newcommand{\Trajeq}{\relax\ifmmode {\mathcal R}_\T \else ${\mathcal R}_\T$\fi} 
\newcommand{\Acteq}{\relax\ifmmode {\mathcal R}_A \else ${\mathcal R}_A$\fi} 
\newcommand{\noop}{\relax\ifmmode \lambda \else $\lambda$\fi} 
\newcommand{\close}[1]{\relax\ifmmode \overline{#1} \else $\overline{#1}$\fi}

\newcommand{\ie}{i.e.,\xspace}

\newcommand{\eg}{e.g.,\xspace}

\newcommand{\tup}[1]
           {
             \relax\ifmmode
             \langle #1 \rangle
             \else $\langle$ #1 $\rangle$ \fi
           }

\newcommand{\lit}[1]{ \relax\ifmmode
                \mathord{\mathcode`\-="702D\sf #1\mathcode`\-="2200}
                \else {\it #1} \fi }

\newcommand{\true}{\relax\ifmmode \mathit true \else \em true \/\fi}

\newlength{\bracklen}

\newcommand{\tri}[3]{\ensuremath{\mathit{#1}^\mathit{#2}_\mathit{#3}}}

\newcommand{\sugLocalVars}[2]{\ifthenelse{\equal{}{#2}}%
                             {\tri{localVars}{#1}{desug}}%
                             {\tri{localVars}{#1}{#2,desug}}}
\newcommand{\sugVars}[2]{\ifthenelse{\equal{}{#2}}%
                        {\tri{vars}{#1}{desug}}%
                        {\tri{vars}{#1}{#2,desug}}}

\newenvironment{subSyntax}{\begin{array}{l}}{\end{array}}

\newcommand{\ms}[1]{\ifmmode%
\mathord{\mathcode`-="702D\it #1\mathcode`\-="2200}\else%
$\mathord{\mathcode`-="702D\it #1\mathcode`\-="2200}$\fi}

\def\T{{\mathcal T}} 

\newcommand{\tuple}[1]{\left\langle#1\right\rangle}

\newcommand{\toolreaffirm}{REAFFIRM\xspace}

\newcommand{\localvar}[2]{{{#1_{#2}}}}

\def\xi{\localvar{x}{i}}

\def\reach{{\sf Reach}}

\def\exec{{\sf Exec}}

\def\Xi{\mathtt{X_i}}

\newcommand{\venc}{v_{enc}}
\newcommand{\nenc}{n_{enc}}
\newcommand{\vgps}{v_{gps}}
\newcommand{\ngps}{n_{gps}}
\newcommand{\nrad}{n_{rad}}
\newcommand{\drad}{d_{rad}}
\newcommand{\dsafe}{d_{safe}}
\newcommand{\vlead}{v_{l}}

\newcommand{\inveg}{{\mathit{inv}}}

\newcommand{\floweg}{{\mathit{flow}}}

\newcommand{\Locset}{{\mathit{Mode}}}

\newcommand{\Transset}{{\mathit{Trans}}}

\newcommand{\src}{\mathit{source}}
\newcommand{\dst}{\mathit{destination}}
\newcommand{\guard}{\mathit{guard}}
\newcommand{\reset}{reset}

\def\Xi{\mathit{X_i}}

\def\AutomatonH{\H}
\newcommand{\Real}{\mathbb{R}}

\def\xi{\localvar{x}{i}}

\def\mode{m}

\begin{document}
\title{REAFFIRM: Model-Based Repair of Hybrid Systems for Improving Resiliency\thanks{This material is based upon work supported by the Defense Advanced Research Projects Agency (DARPA) and Space and Naval Warfare Systems Center Pacific (SSC Pacific) under Contract No. N6600118C4007, the National Science Foundation (NSF) GRFP under Grant No. DGE-1845298, and sponsored in part by ONR N000141712012.}}
%


\author{Luan Viet Nguyen\and
Gautam Mohan\and
James Weimer\and Oleg Sokolsky\and Insup Lee\and Rajeev Alur}
\authorrunning{Luan Nguyen et al.}
\institute{Department of Computer and Information Science, University of Pennsylvania, PA, USA}
%
\maketitle              
%
%
\begin{abstract}
\label{sec:abstract}
Model-based design offers a promising approach for assisting developers to build reliable and secure cyber-physical systems (CPSs) in a systematic manner. In this methodology, a designer first constructs a model, with mathematically precise semantics, of the system under design, and performs extensive analysis with respect to correctness requirements before generating the implementation from the model.
However, as new vulnerabilities are discovered, requirements evolve aimed at ensuring resiliency. 
There is currently a shortage of an inexpensive, automated mechanism that can effectively repair the initial design, and a model-based system developer regularly needs to redesign and reimplement the system from scratch.

In this paper, we propose a new methodology along with a MATLAB toolkit called \toolreaffirm to facilitate the model-based repair for improving the resiliency of CPSs.  
%
\toolreaffirm takes as inputs 1) an original hybrid system modeled as a Simulink/Stateflow diagram, 2) a given resiliency pattern specified as a model transformation script, and 3) a safety requirement expressed as a Signal Temporal Logic formula, and outputs a repaired model which satisfies the requirement. 
The overall structure of \toolreaffirm contains two main modules: a model transformation and a model synthesizer built on top of the falsification tool Breach.  
We also introduce a new model transformation language for hybrid systems, which we call HATL to allow a designer to specify resiliency patterns. 
To evaluate the proposed approach, we use \toolreaffirm to automatically synthesize repaired models for an adaptive cruise control (ACC) system under a GPS sensor spoofing attack, for a single-machine infinite-bus (SMIB) system under a sliding-mode switching attack, and for a missile guidance system under gyroscopes sensor attack.
\end{abstract} 		
\section{Introduction}
\label{sec:intro}
A cyber-physical system (CPS) consists of computing devices communicating with one another and interacting with the physical world via sensors and actuators. Increasingly, such systems are everywhere, from smart buildings to autonomous vehicles to mission-critical military systems. 
The rapidly expanding field of CPSs precipitated a corresponding growth in security concerns for these systems. The increasing amount of software, communication channels, sensors and actuators embedded in modern CPSs make them likely to be more vulnerable to both cyber-based and physics-based attacks~\cite{wan2015security,wasicek2014aspect,kocher2004security,al2015design,gamage2010enforcing}. As an example, \emph{sensor spoofing} attacks to CPSs become prominent, where a hacker can arbitrarily manipulate the sensor measurements to compromise secure information or to drive the system toward unsafe behaviors. Such attacks have successfully disrupted the braking function of the anti-lock braking systems~\cite{Shoukry2013,al2015design}, and compromised the insulin delivery service of a diabetes therapy system~\cite{li2011hijacking}. Alternatively, attackers can gain access to communication channels to either manipulate the switching behavior of a smart power grid~\cite{liu2011class} or disable the brake system of a modern vehicle~\cite{koscher2010experimental}. Generally, constructing a behavioral model at design time that offers resiliency for all kinds of attacks and failures is notoriously difficult. 
\begin{figure}[t!]%
	\centering
		\includegraphics[width=0.8\textwidth]{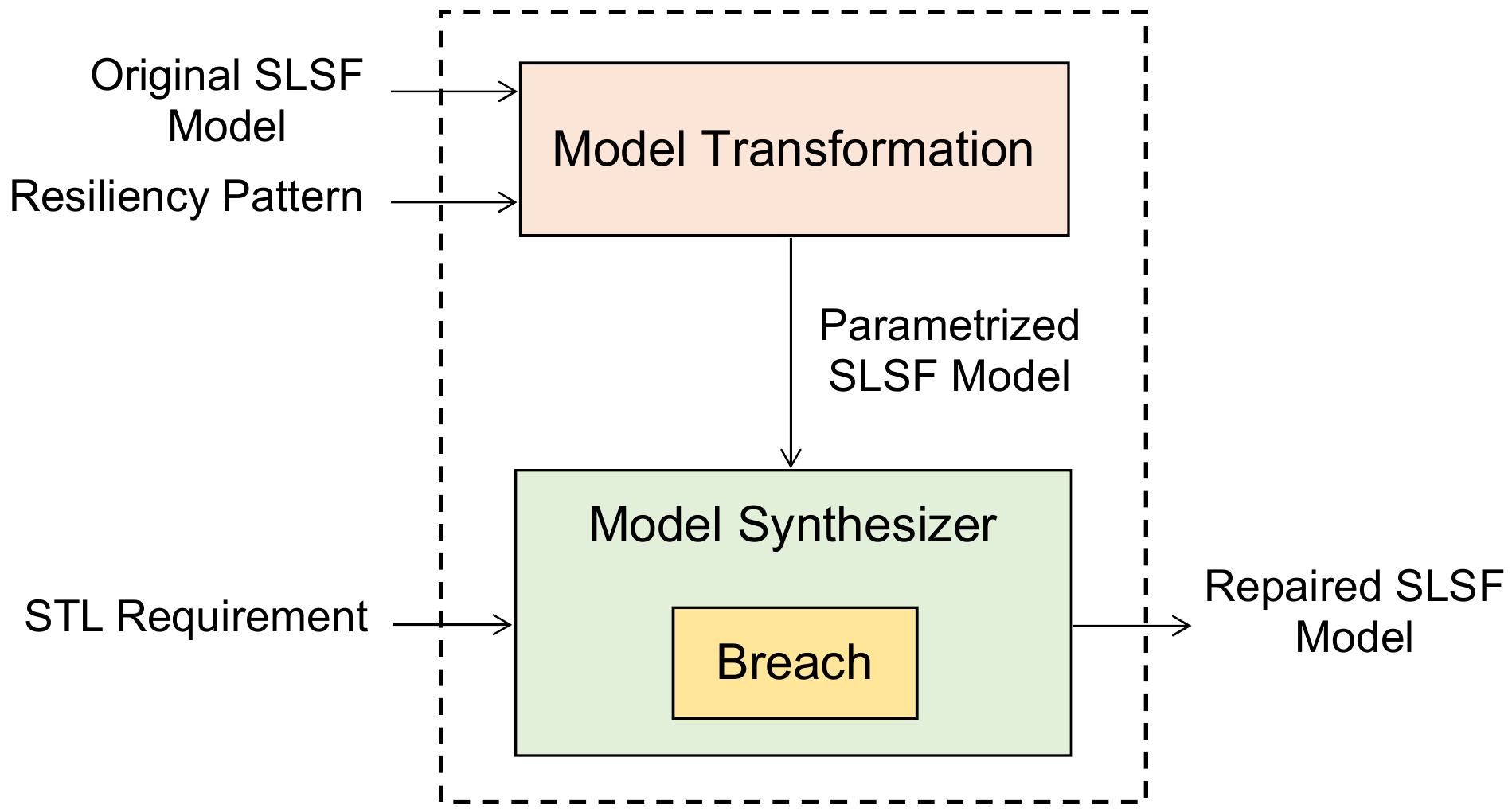}%
	\caption{\toolreaffirm overview.} 
	\figlabel{overview}%
\end{figure}%
Traditionally a model of a CPS consists of block diagrams describing the system architecture and a combination of state machines and differential equations describing the system dynamics~\cite{alur1995algorithmic}. Suppose a designer has initially constructed a model of a CPS that satisfies correctness requirements, but at a later stage, this correctness guarantee is invalidated, possibly due to adversarial attacks on sensors, or violation of environment assumptions. Current techniques for secure-by-design systems engineering do not provide a formal way for a designer to specify a resiliency pattern to automatically repair system models based on evolving resiliency requirements under unanticipated attacks. 

In this paper, we propose a new methodology and an associated toolkit, called \toolreaffirm, to assist a designer in repairing the original model so that it continues to satisfy the correctness requirements under the modified assumptions. 
The proposed technique relies on designing a collection of \emph{potential edits} (or \emph{resiliency patterns}) to the original model to generate the new model whose parameters values can be determined by solving the \emph{parameter synthesis problem}. 
\figref{overview} shows an overview of \toolreaffirm, which contains two mains modules 1) a \emph{model transformation}, and 2) a \emph{model synthesizer} built on top of the falsification tool Breach~\cite{donze2010breach}.
\toolreaffirm takes the following inputs 1) the original system modeled as a Simulink/Stateflow (SLSF) diagram, 2) the resiliency pattern specified by the designer and 3) the safety requirement expressed as a Signal Temporal Logic (STL)~\cite{maler2004monitoring} formula, and outputs the repaired SLSF model that satisfies the safety requirement.

To allow a designer to specify resiliency patterns we have developed a new \emph{model transformation language} for hybrid systems, called HATL (Hybrid Automata Transformation Language).
A HATL script is a sequence of statements that describe the modifications over the structure of hybrid systems modeled as hybrid automata~\cite{alur1995algorithmic}.  Examples of edits to a model include creating new modes of operations, duplicating modes, adding transitions, modifying switching conditions, and substituting state variables in flow equations. 
The proposed language allows the designer to write a resiliency pattern in a generic manner, and programmatically modify the initial design without knowing the internal structures of a system. The HATL interpreter is implemented in Python with an extensible backend to allow interoperability with different hybrid systems modeling frameworks. The current implementation of HATL supports MATLAB and performs transformations on SLSF models.

For evaluation, we apply \toolreaffirm to automatically synthesize the repaired models for three case studies in the domains of automotive control, smart power systems and aerospace applications. The first case study is a simplified model of an adaptive cruise control (ACC) system under a GPS sensor spoofing attack, and the resiliency pattern to fix the model is to ignore the GPS measurement and only use the wheel encoders, which are additional (redundant) sensors for estimating a vehicle's velocity. \toolreaffirm automatically synthesizes the condition that triggers a switch to a copy of the model that ignores the GPS measurement. The second case study is a single-machine infinite-bus (SMIB) model, which is an approximation of a smart power grid, under a sliding-mode attack. In this case, the mitigation strategy is to increase the minimal dwell-time to avoid rapid changes between different operation modes. Thus, the resiliency pattern adds a dwell-time variable in each mode of the model, and the minimal dwell-time can be synthesized automatically by \toolreaffirm. The third case study is the missile guidance system (MG) provided by Mathworks, which is a good representative of a practical MG system as it has more than 300 SLSF blocks. The principle of a spoofing attack on the gyroscopes of the MG system is similar to the GPS spoofing attack of the ACC system, and then we can apply the same resiliency pattern used to fix the ACC model for repairing the MG model.

In summary, the main contributions of the paper are as follows.
\begin{enumerate}[leftmargin= 1em]
\item The methodology to facilitate the model-based repair for improving the resiliency of CPSs against unanticipated attacks and failures,
\item the design and implementation of an extensible model transformation language for specifying resiliency patterns used to repair CPS models,
\item the end-to-end design and implementation of the toolkit, which integrates the model transformation and the model synthesis tools to automatically repair CPS models,
\item the applicability of our approach on three proof-of-concept case studies where the CPS models can be repaired to mitigate practical attacks.
\end{enumerate}

The remainder of the paper is organized as follows.~\secref{overview} presents an overview of our proposed methodology through a simplified example of the ACC system, and introduces the architecture of \toolreaffirm.~\secref{transformation} describes our model transformation language used to design a resiliency pattern for hybrid systems.~\secref{synthesis} presents the model synthesizer of \toolreaffirm.~\secref{result} presents three case studies that illustrate the capability of \toolreaffirm in automatically repairing the original models of a) the ACC system under a GPS sensor spoofing attack, b) the SMIB system under a sliding-mode attack, and c) the MG system under a gyroscopes sensor attack.~\secref{rw} reviews the related works to \toolreaffirm, and ~\secref{conclude} concludes the paper.
%

\section{Overview of the Methodology}
\seclabel{overview}
%
%
%
In this section, we will explain our methodology through a simplified example of the  adaptive cruise control (ACC) system. 
Assume that a designer has previously modeled the ACC system as a combination of the vehicle dynamics and an ACC module, and GPS measurements were considered trusted in the initial design. In the following, we will describe the ACC system as originally designed, an attack scenario, and an example of resiliency pattern to repair the ACC model\footnote{We note that the ACC model presented herein is not a representative of the complexity of a true ACC system, but a simplified example in which the dynamics and control equations are chosen for simplicity of presentation.}.
Then, we present how \toolreaffirm can automatically perform a model transformation and synthesis to construct a new ACC model with resiliency. 
%
%
\subsection{A Simplified Example of ACC System}
%
%
For simplicity, we assume that the designer initially models the ACC system (including vehicle dynamics) as a hybrid system shown in \figref{original}.
The original ACC system operates in two modes: \emph{speed control} and \emph{spacing control}. In speed control, the host car travels at a driver-set speed. In spacing control, the host car aims to maintain a safe distance from the lead car. 
The vehicle has two state variables: $d$ is the distance to the lead car, and $v$ is the speed of the host vehicle. The ACC system has two sensors that measure its velocity $v$ via noisy wheel encoders, $\venc = v + \nenc$, and a noisy GPS sensor, $\vgps = v+\ngps$, where $\nenc$ and $\ngps$ denote the encoder and GPS noises, respectively. Additionally, the ACC system has a radar sensor that measures the distance to the lead vehicle, $\drad = d+\nrad$, where $\nrad$ captures a corresponding noise.
The ACC system decides which mode to use based on the real-time sensor measurements. For example, if the lead car is too close, the controller triggers the transition $g_{sd}$ to switch from speed control to spacing control. Similarly, if the lead car is further away, the ACC system switches from spacing control to speed control by executing the transition $g_{ds}$. 
The \emph{safety specification} of the system is that $d$ should always be greater than $\dsafe$, where $\dsafe = v + 5$. We will describe the ACC model in more details in~\secref{result}.
%
%
%
%
%
%
\begin{figure}%
	\centering%
		\includegraphics[width=0.7\textwidth]{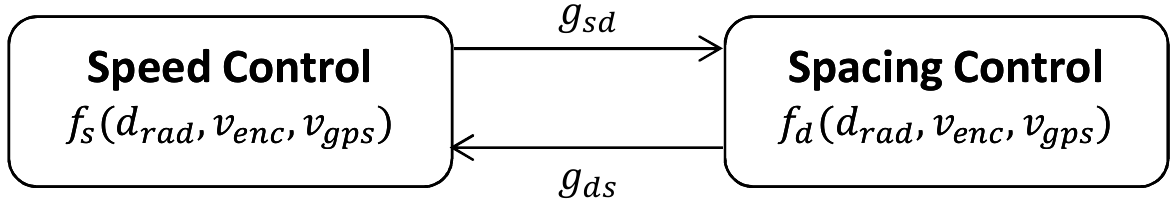}%
	\caption{An original ACC model.}%
	\figlabel{original}%
	\vspace{-1em}
\end{figure}%
%

\noindent
{\bf Safety violation under GPS sensor attack.}
In this example, we assume that after designing and verifying the initial ACC system, it is determined that the GPS sensor can be \emph{spoofed} \cite{tippenhauer2011requirements,kerns2014unmanned}. GPS spoofing occurs when incorrect GPS packets (possibly sent by a malicious attacker) are received by the GPS receiver. In the ACC system, this allows an attacker to arbitrarily change the GPS velocity measurement. 
Thus, a new scenario occurs when the original assumption of GPS noise, \eg $|\ngps| \leq 0.05$ is omitted, and the new assumption is $|\ngps| \leq 50$.
As a result, the safety specification could be violated under the GPS sensor attacks, and a designer needs to repair the original model using a known mitigation strategy. 
\begin{figure}%
	\centering%
		\includegraphics[width=0.7\textwidth]{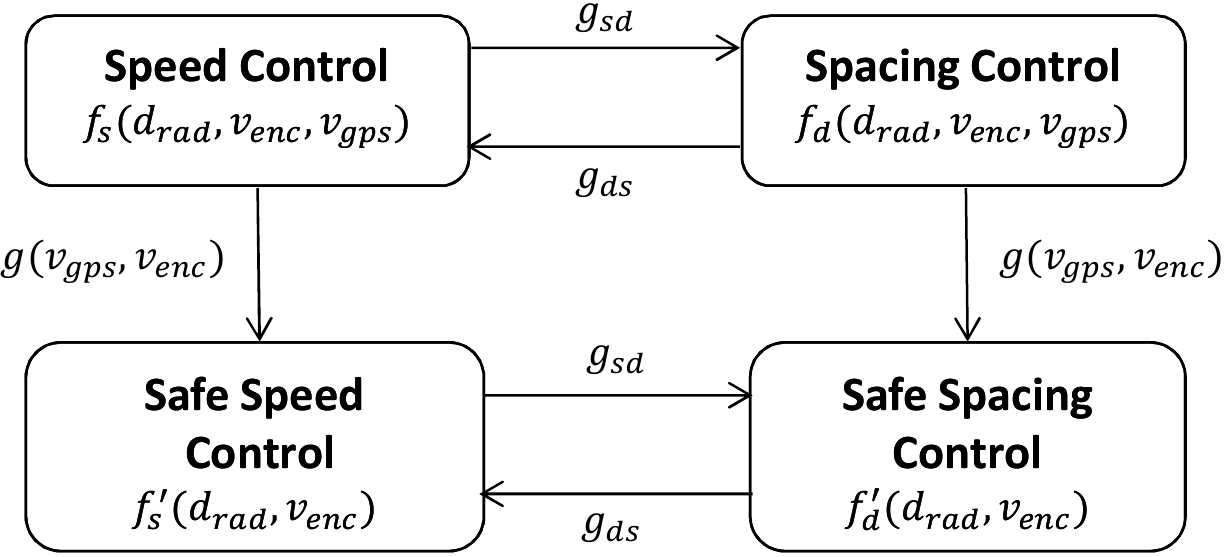}%
	\caption{A repaired ACC model without a reference to GPS sensor under spoofing attacks.}%
	\figlabel{updated}%
	\vspace{-1em}
\end{figure}%
%

\noindent
{\bf Example of resiliency pattern: ignoring GPS measurement.}
Since the ACC system has redundancy in the sensory information of its estimated velocity, to provide resilience against the GPS attacks, a mitigation strategy is to ignore the GPS value, and use only the wheel encoders to estimate velocity. 
Thus, a potential fix is first to create a copy of the original model where the controller simply ignores the GPS reading as it can no longer be trusted. 
%
Then, adding new transitions from the modes of the original model to the corresponding instances of the copy that uses only the wheel encoder to measure velocity.
We note that this transformation is generic, that is, it can be applied in a uniform manner to any given model simply by creating a duplicate version of each original mode and transition, copying the dynamics in each mode, but without a reference to the variable $\vgps$. 

\figref{updated} illustrates the repaired model in which the transition from the original speeding and spacing control modes to their copies is an expression over $\vgps$ and $\venc$. 
Observe that while it would be possible to use only the wheel encoders all the time, a better velocity estimate can be obtained by using an average velocity measurement (from both the GPS and wheel encoders) when the GPS sensor is performing within nominal specifications. The main analysis question is when should the model switch from the original modes to the copied modes during the spoofing attack. 
From a practical standpoint, such a transition should occur when the GPS measurement significantly deviates from the wheel encoder measurement, and a transition condition can be specified as $g(\vgps,\venc) = |\vgps -\venc| \geq \theta$, where $\theta$ is an unknown parameter. 
Since $\venc = v + \nenc$ and $\vgps = v+\ngps$, we can rewrite the transition condition as $g(\vgps,\venc) = |\ngps -\nenc| \geq \theta$.
Then, one needs to synthesize the suitable value of the parameter $\theta$ that specifies the threshold for switching from the original copy to the new copy so that the safety requirement is satisfied.

\subsection{REAFFIRM Toolkit}
Our \toolreaffirm prototype for the model-based repair is built in MATLAB and consists of two main modules, corresponding to \emph{model transformation} and \emph{parameter synthesis}. To synthesize the model with resiliency to unanticipated attacks, users need to provide the following inputs to \toolreaffirm:
\begin{itemize}[leftmargin= 1 em]
    \item the initial design of a hybrid system modeled in MathWorks SLSF format,
		\item the resiliency pattern specified as a model transformation script that transforms the initial model to the new model with resiliency to unanticipated attacks, and
		\item the correctness (safety) requirement of the system specified as an STL formula.
%
\end{itemize}

In the case of the ACC example, the inputs of \toolreaffirm are the initial SLSF model shown in~\figref{original}, the resiliency pattern that creates the copied version of the original model without a reference to the variable $\vgps$, and the safety requirement encoded as an STL formula,
\begin{align}
\varphi_{ACC} &= \Box_{[0, \infty)} d[t] < 5 + v[t]. 
\end{align}
The model transformation tool of \toolreaffirm takes the initial SLSF model and the resiliency pattern (\eg the transformation script shown in~\figref{examplecode}), and then generates the new SLSF model that contains a parameter $\theta$ that appears in the switching condition based on the difference between the GPS measurement and the wheel encoder measurement.
Then, the model synthesizer tool of \toolreaffirm takes the parametrized ACC model in SLSF and the STL formula $\varphi_{ACC}$ as inputs, and then performs a parameter synthesis to find the desired value of $\theta$ over a certain range, to ensure that $\varphi_{ACC}$ is satisfied. Internally, the model synthesizer of \toolreaffirm utilizes an open-source model falsification tool---Breach~\cite{donze2010breach} to synthesize the desired parameters values. If the synthesizer can find the best value of $\theta$ over the given range, then \toolreaffirm outputs a competed SLSF model which satisfies $\varphi_{ACC}$ under the GPS attacks. Otherwise, the tool will suggest the designer to either search over different parameter ranges or try different resiliency patterns to repair the ACC model.   
\section{Model Transformation}
\seclabel{transformation}
%
\subsection{Representation of Hybrid System}
Hybrid automata~\cite{alur1995algorithmic} are a modeling formalism popularly used to model hybrid systems which include both continuous dynamics and discrete state transitions. A hybrid automaton is essentially a finite state machine extended with a set of real-valued variables evolving continuously over time~\cite{alur1995algorithmic}. 
The main structure of a hybrid automaton $\AutomatonH$ includes the following components.
%
%
%
%
%

%
\begin{itemize}[leftmargin= 1em]
%
\item $\X$: the finite set of $n$ continuous, real-valued variables. 
\item $\P$: the finite set of $p$ real-valued parameters.
\item $\Locset$: the finite set of discrete modes. For each mode $\mode \in \Locset$, $\mode.\inveg$ is an expression over $\X \cup \P$ that denotes the invariant of mode $\mode$, and $\mode.\floweg$ describes the continuous dynamics governed by a set of ordinary differential equations. 
\item $\Transset$: the finite set of transitions between modes. Each transition is a tuple $\tau \deq \tuple{\src, \dst, \guard, \reset}$, where $\src$ is a source mode and $\dst$ is a target mode that may be taken when a guard condition $\guard$, which is an expression over $\X \cup \P$, is satisfied, and $\reset$ is an assignment of variables in $\X$ after the transition. 
%
%
\end{itemize}
We use the dot (.) notation to refer to different components of tuples, \eg $\AutomatonH.\Transset$ refers to the transitions of automaton $\AutomatonH$ and $\tau.\guard$ refers to the guard of a transition $\tau$. Since our goal is to repair a hybrid automaton syntactically, we will not discuss its semantics in this paper, but refer a reader to~\cite{alur1995algorithmic} for details. 
We note that the \emph{model transformation language} proposed in this paper transforms a hybrid automaton based on modifying the syntactic components of the hybrid automaton in a generic manner. 
The transformation tool of \toolreaffirm can take a HATL transformation script and translate it into an equivalent script that performs a model transformation for different modeling framework of hybrid automata including a continuous-time Stateflow chart.

\vspace{0.5em}
\noindent
{\bf Continuous-time Stateflow chart.}
%
In this paper, we represent hybrid automata using \emph{continuous-time} Stateflow chart, which is a standard commercial modeling language for hybrid systems integrated within Simulink.
A continuous-time Stateflow chart supplies methods for engineers to quickly model as well as efficiently refine, test, and generate code for hybrid automata.
The syntactic description of a continuous-time Stateflow chart is basically a hybrid automaton, with a small few differences. In particular, a mode is a \emph{state} associated with different types of actions including a) \emph{entry} action executed when entering the state, b) \emph{exit} executed when exiting the state, and c) \emph{during} (or \emph{du}) action demonstrates the continuous-time evolution of the variables (\ie $\floweg$ dynamics) when no transition is enabled. A variable can be specified as \emph{parameter}, \emph{input}, \emph{output}, and \emph{local variable}. Also, an SLSF model which includes a continuous-time Stateflow chart is deterministic since its transition is urgent and executed with priorities~\cite{bak2017hybrid}.
%

\subsection{Hybrid Automata Transformation Language}
\begin{figure}[!t]
\begin{lstlisting}[basicstyle=\ttfamily\scriptsize, numbers=none]
# original model is retrieved from command line arguments
model_copy = model.copyModel() # make a model copy 
# start a transformation  
model.addParam("theta") # add new parameter theta
formode m = model.Mode {
		m_copy = model.addMode(m)
		m.replace(m_copy.flow,"ngps","nenc")
		model.addTransition(m,m_copy,"abs(ngps-nenc)>theta")
}
fortran t = model_copy.Trans {
		# get source and destination modes of transition t
		src = t.source
		dst = t.destination
		# retrieve copies of source and destination modes
		src_copy = model.getCopyMode(src)
		dst_copy = model.getCopyMode(dst)
		model.addTransition(src_copy,dst_copy,t.guard)
}
# end of the transformation
\end{lstlisting}
\caption{An example of a resiliency pattern written as a HATL script for the ACC system.}%
\figlabel{examplecode}%
\end{figure}

In our approach, the partial model of the system, which satisfies functional but not necessarily resiliency requirements is originally modeled in the form of hybrid automata. The model transformation that is at the core of the \toolreaffirm tool will then attempt to modify the components of the automata such as modes, flows, or switching logic, by applying user-defined resiliency patterns.

In order to specify resiliency patterns for hybrid automata, we introduce a new language for model transformation called HATL (Hybrid Automata Transformation Language). The goal of HATL is to allow a designer to repair an original model in a programmatic fashion. HATL scripts abstract model implementation details so engineers do not need to learn the intricacies of an individual framework. A key use of HATL is to write generic scripts that are applicable to many models, promoting resiliency scripts which are reusable.
A script written in HATL is a sequence of \emph{statements} that specify the changes over the structure of given hybrid automata. HATL's syntax and semantics are designed to make it intuitive to anyone who is familiar with imperative languages. HATL includes \emph{loop} statements that iterate over sets of objects, such as modes or transitions of a model. It uses \emph{dot references} to index into structures to obtain data fields or to call object-specific methods. \emph{Assignments} are mutable, and scoped within statement blocks. Functions and methods can have variable numbers of arguments which are eagerly evaluated.

The model transformation tool built in \toolreaffirm takes a resiliency pattern in the form of a HATL script, and then translates each of the statements of the script into equivalent transformation operations on continuous-time Stateflow models.
%
~\figref{examplecode} shows an example of a HATL script that specifies a transformation from the original ACC model shown in~\figref{original} to the parametrized model shown in~\figref{updated}. 
In this script, we first create a copy of the original model. Next, we iterate over each mode of the model by calling the \emph{formode loop}, make a copy with replacing the variable $\ngps$ by $\nenc$, and then add a new transition from the original mode to the copied mode with a new guard condition. This guard condition is a constraint specified over the difference between $\ngps$ by $\nenc$ (\ie the difference between $\vgps$ and $\venc$)  and a new parameter $\theta$, which is added into the model using a function call \emph{addParam}. 
Finally, we need to copy all transitions between original modes (stored in a copied version of the original model) and assign them to the corresponding duplicated modes.

%
\subsection{Implementation}
%
Our current implementation dynamically interprets HATL scripts in
Python and translates them into SLSF model transformations via
the MATLAB Engine. Our interpreter checks argument values at runtime
to ensure only valid transformed models are produced. If a malformed
program statement is detected, HATL will throw a verbose error message
and roll back any changes it has applied already before
exiting. Additionally, these error messages are reported in terms of
generic HATL models, so an engineer writing a resiliency pattern does
not need to worry about the underlying implementation.

Currently, HATL provides enough programming abstraction to express concise model transformations that function as valid resiliency patterns, and more examples of these scripts will be introduced in \secref{result}. There is room for future improvement, such as adding language constructs like type checking to verify the type correctness of the model before and after repair. 

\section{Model Synthesis}
\seclabel{synthesis}
In this section, we present the model synthesizer incorporated in \toolreaffirm which takes a parameterized model produced by the model transformation, and a correctness requirement as inputs, and then generates a completed model with parameter values instantiated to satisfy the correctness requirements. Since the structure of the completed model is already determined after the model transformation, the model synthesis problem then reduces to the \emph{parameter synthesis problem}. Let $\P_s$ be the set of parameters of the transformed model $\tilde{\AutomatonH}$, given a safety specification $\varphi$ and sets of parameter values $\bar{\P_s}$ , find the best instance values of $\P_s$ over $\bar{\P_s}$ so that $\tilde{\AutomatonH} \models\varphi$.  
For example, the transformation of the ACC model shown in \figref{updated} introduces a new parameter $\theta$ whose value needed to be determined so that the completed model will satisfy the safety requirement with respect to the same initial condition of the state variables and parameters domains of the original model.
%
%
\subsection{Overview of Breach}
%
%
We incorporated Breach into the model synthesizer of \toolreaffirm as an analysis mechanism to perform the falsification and parameter synthesis for hybrid systems. Given a hybrid system modeled as an SLSF diagram, an STL specification described the safety property, and specific parameter domains, Breach~\cite{donze2010breach} can perform an optimized search over the parameter ranges to find parameter values that cause the system violating the given STL specification. 
The parameter mining procedure is guided by the counterexample obtained from the falsification, and it terminates if there is no counterexample found by the falsifier or the maximum number of iterations specified by a user is reached.
On the other hand, Breach can compute the sensitivity of execution traces to the initial conditions, which can be used to obtain completeness results by performing systematic simulations. Moreover, Breach provides an input generator for engineers to specify different testing input patterns such as step, pulse width, sinusoid, and ramp signals. This input generator is designed to be extensible, so users can write a specific input pattern to test their model against particular attack scenarios.

We note that although Breach cannot completely prove the system correctness, it can efficiently find bugs existing in the initial design of CPS that are too complex to be formally verified~\cite{kapinski2015simulation}. These bugs are essential for an engineer to specify resiliency patterns to repair the model. 
Moreover, the general problem of verifying a CPS modeled as a hybrid system is known to be \emph{undecidable}~\cite{henzinger1995s}. 
Instead, the falsification algorithms embedded within Breach are scalable and work properly for black-box hybrid systems with different classes of dynamics.
Thus, in practice, engineers prefer to use counterexamples obtained by a falsification tool to refine their design. Our prototype \toolreaffirm utilizes the advantages of SLSF modeling framework and the falsification tool Breach to design a resiliency pattern and perform the model synthesis for a repaired CPS model with resiliency.
%
\subsection{Model Synthesis using Breach}
Next, we describe how REAFFIRM uses Breach to synthesize parameters values for the parametrized model returned from the model transformation tool. The parameter synthesis procedure consists of following steps.
\begin{enumerate}[leftmargin= 1em]
\item We first specify the initial conditions of state variables and parameters, the set of parameters $\P_s$ that need to be mined, the sets of parameter values $\bar{\P_s}$, and the maximum time (or number of iterations) for the optimization solver of Breach.
\item Next, we call the falsification loop within Breach to search for a counterexample. For each iteration, if the counterexample is exposed, the unsafe values of $\P_s$ will be returned. Based on these values, the tool will automatically update the sets of parameter values $\bar{\P_s}$ to the new sets of parameter values $\bar{\P'_s} \subset \bar{\P_s}$, and then continue the falsification loop.
\item The process repeats until the property is satisfied that means the falsifier cannot find a counterexample and the user-specified limit on the number of optimized iterations (or time) for the solver expires.  
\item Finally, the tool returns the best (and safe) values of $\P_s$, updates the parametrized model with these values, and then exports the completed model. If the synthesizer fails to find the values of $\P_s$ over the given sets of parameter values $\bar{\P_s}$ so that the safety requirement is satisfied, it will recommend a designer to either search over different parameter ranges or try another resiliency pattern.
\end{enumerate}
\vspace{0.5em}
{\bf Monotonic Parameters.} The search over the parameter space of the synthesis procedure can be significantly reduced if the satisfaction value of a given property is monotonic w.r.t to a parameter value. Intuitively, the satisfaction of the formula monotonically increases (respectively decreases) w.r.t to a parameter $p$ that means the system is more likely to satisfy the formula if the value of $p$ is increased (respectively decreased). In the case of monotonicity, the parameter space can be efficiently truncated to find the \emph{tightest} parameter values such that a given formula is satisfied. In Breach, the check of monotonicity of a given formula w.r.t specific parameter is encoded as an STM query and then is determined using an STM solver. However, the result may be \emph{undecidable} due to the undecidability of STL~\cite{jin2015mining}. 
In this paper, the synthesis procedure is based on the assumption of satisfaction monotonicity. If the check of monotonicity is undecidable over a certain parameter range, a user can manually enforce the solver with decided monotonicity (increasing or decreasing) or perform a search over a different parameter range.

\section{Model Repair for Resiliency}
\seclabel{result}
%
%
In this section, we demonstrate the capability of \toolreaffirm to repair CPSs models under unanticipated attacks. We first revisit the ACC example and evaluate three resiliency patterns that can be applied to repair the ACC model under the GPS sensor spoofing attack. Second, we investigate a sliding-mode switching attack that causes instability for a smart grid system and how \toolreaffirm can use a dwell-time pattern to repair the model under this attack automatically. Finally, we apply the three resiliency patterns used to fix the ACC model to repair the MG system under gyroscopes sensor attacks.

\toolreaffirm was tested using MATLAB 2018a and MATLAB 2018b executed on an x86-64 laptop with 2.8 GHz Intel(R) Core(TM) i7-7700HQ processor and 32 GB RAM. All performance metrics reported were recorded on this system using MATLAB 2018a. In Breach, we choose the CMAES solver, and the maximum optimization time is 30 seconds for each iteration of the falsification loop. \toolreaffirm and all case studies investigated in this paper are available to download at \url{https://github.com/LuanVietNguyen/reaffirm}.
The overall performance\footnote{The transformation time reported in the table is the actual time required for the model transformation by neglecting the overhead of loading the MATLAB Engine in Python.} of \toolreaffirm in repairing the initial models of three case studies to mitigate their corresponding attacks is summarized in \tabref{performance_results}. Next, we will describe three case studies in more details.

\begin{table}[t!]
\large
\centering
\resizebox{0.98\linewidth}{!}{\begin{tabular}{|l|l|l|l|l|l|l|l|l|l|} 
\hline
\textbf{Model}       & \textbf{BD}	&\textbf{Attack}                                                 & \multicolumn{2}{l|}{\textbf{Resiliency Pattern }}                                                                                                                          & \textbf{Unknown Condition}                                                               & \textbf{PR}              & \textbf{SV} & \textbf{TT} & \textbf{ST}  \\ 
\hline
\multirow{3}{*}{ACC} & \multirow{3}{*}{11} & \multirow{3}{*}{GPS spoofing}                                   & \multirow{3}{*}{\begin{tabular}[c]{@{}l@{}}Ignore GPS,~\\ measurement, use~\\ wheel~encoders value\end{tabular}} & Pattern 1                                                    & \multirow{2}{*}{\begin{tabular}[c]{@{}l@{}}When to switch to \\a safe copy\end{tabular}} & \multirow{2}{*}{$\theta \in [0, 50]$} & 7.08515~~                  & 2                    & 88              \\ 
\cline{5-5}\cline{8-10}
                     &                                                                 &                                                                                         &                    & Pattern 2                                                    &                                                                                          &                                       & 7.08515~ ~~                & 2                    & 88             \\ 
\cline{5-10}
                     &                                                                 &                                                                                        &                     & Pattern 3                                                    & \begin{tabular}[c]{@{}l@{}}Ratio of GPS/encoders\\measurements\end{tabular}              & $\theta \in [0.1, 0.9]$               & 0.1543~                    & 1.75                & 55.78            \\ 
\hline
SMIB                 & 15 & \begin{tabular}[c]{@{}l@{}}Sliding-mode\\switching\end{tabular} & \begin{tabular}[c]{@{}l@{}}Add a dwell-time to\\ avoid rapid switching\end{tabular}                    & \begin{tabular}[c]{@{}l@{}}Pattern \\dwell-time\end{tabular} & Minimal dwell-time                                                                       & $\theta \in [0, 0.3]$                 & 0.12                       & 2                   & 45               \\
\hline
\multirow{3}{*}{MG} & \multirow{3}{*}{310} & \multirow{3}{*}{\begin{tabular}[c]{@{}l@{}}Gyroscopes\\spoofing\end{tabular}}                                   & \multirow{3}{*}{\begin{tabular}[c]{@{}l@{}}Ignore untrusted,~\\ measurements, use~\\ the trusted ones\end{tabular}} & Pattern 1                                                    & \multirow{2}{*}{\begin{tabular}[c]{@{}l@{}}When to switch to \\a safe copy\end{tabular}} & \multirow{2}{*}{$\theta \in [0, 0.5]$} & 0.06714~~                  & 2                    & 78              \\ 
\cline{5-5}\cline{8-10}
                     &                                                                 &                                                                                       &                      & Pattern 2                                                    &                                                                                          &                                       & 0.06714~ ~~                & 2                    & 92            \\ 
\cline{5-10}
                     &                                                                 &                                                                                       &                      & Pattern 3                                                    & \begin{tabular}[c]{@{}l@{}}Ratio of gyroscopes\\ measurements\end{tabular}              & $\theta \in [0.01, 0.1]$               & 0.01127                    & 1.75                & 55           \\ 
\hline
\end{tabular}}
\vspace{1em}
\caption{REAFFIRM performance results for the ACC, SMIB, and MG case studies. BD is the number of blocks in SLSF models. PR is the parameter range. SV is the synthesized value. TT is the transformation time (s). ST is the synthesis time (s). }
\label{tab:performance_results}
\end{table}
\subsection{Adaptive Cruise Control System}

\noindent
{\bf Original SLSF model.} We previously introduced the simplified example of the ACC system in~\secref{overview} to illustrate our approach. In this section, we present the ACC system in more details. The ACC system can be modeled as the SLSF model shown in \figref{acc_slsf_model}. 
The model has four state variables where $d$ and $e_d$ are the actual distance and estimated distance between the host car and the lead vehicle, $v$ and $e_v$ represent the actual velocity and estimated velocity of the host car, respectively.  
In this model, we assume that the lead vehicle travels with a constant speed $\vlead$. The transition from speed control to spacing control occurs when the estimate of the distance is less than twice the estimated safe distance, \ie $e_d < 10 + 2e_v$. A similar condition is provided for switching from spacing control to speed control, \ie $e_d \geq 10 + 2e_v$. In this case study, we assume that the designer has verified the initial SLSF model of the ACC system against the safety requirement $\varphi_{ACC}$ under the scenario when $d(0) \in [90, 100]$, $v(0) \in [25, 30]$, $|d(0) - e_d(0)| \leq 10$, $|v(0) - e_v(0)| \leq 5$ , $\vlead  = 20$, $|\nrad| \leq 0.05$, $|\nenc| \leq 0.05$ and $|\ngps| \leq 0.05$.


\begin{figure}[t!]%
	\centering%
	\vspace{0.5em}
    \includegraphics[width=0.8\textwidth]{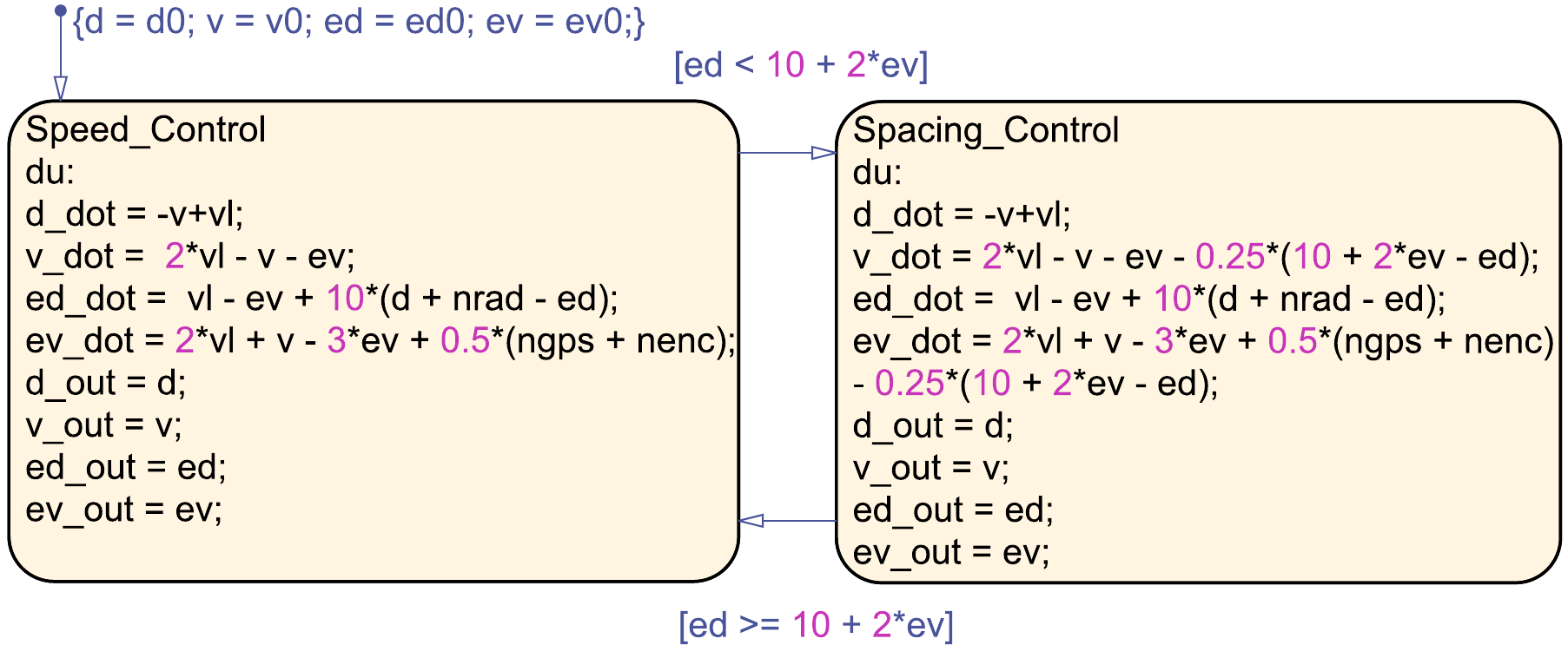}
	\caption{The original SLSF model of the ACC system.}%
	\figlabel{acc_slsf_model}%
\end{figure}%

\vspace{0.5em}
\noindent
{\bf GPS sensor attack.} To perform a spoofing attack on the GPS sensor of the  ACC model, we continuously inject false data to manipulate its measurement value. In this case, we omit the original assumption $|\ngps| \leq 0.05$, and employ the new assumption as $|\ngps| \leq 50$. Using the input generator in Breach, we can specify the GPS spoofing attack as a standard input test signal such as a constant, ramp, step, sinusoid or random signal. The following evaluations of three different resiliency patterns used to repair the ACC model are based on the same assumption that the GPS spoofing occurs at every time point, specified as a random constant signal over the range of [-50, 50] during 50 seconds. 

\vspace{0.5em}
\noindent
{\bf Model Repair for the ACC system.} Under the GPS sensor spoofing  attack, the original SLSF model does not satisfy its safety requirement and a designer needs to apply a certain resiliency pattern to repair the model. The \emph{first resiliency pattern} for repairing the ACC system has been introduced in~\secref{overview}, which makes the copy of the original model where the controller ignores the GPS reading as it can no longer be trusted. However, we need to determine the best switching condition from the original model to the copy.~\figref{acc_model_pat1} in \appref{additional} shows the completed model, where the switching condition is determined by synthesizing the value of $\theta$ over the range of [0, 50] using Breach. 

The \emph{second resiliency pattern} for the ACC model is the extended version of the first one where it includes a switching-back condition from the copy to the original model when the GPS sensor attack is detected and mitigated. An example of such a switching-back condition is when the difference between the $\nenc$ and $\ngps$ are getting smaller, \ie $|\ngps -\nenc| < \theta - \epsilon$, where $\epsilon$ is a positive user-defined tolerance. For this pattern, the model transformation script can be written similar to the one shown in~\figref{examplecode} with adding the \emph{addTransition} function from the copy mode to the original mode with the guard condition labeled as $|\ngps -\nenc| < \theta - \epsilon$ in the \emph{formode} loop.
The performance of \toolreaffirm for the second pattern is similar to the first pattern with the same synthesized value of $\theta = 7.08515$ and $\epsilon = 0$.

\begin{figure}[t!]
\begin{lstlisting}[basicstyle=\ttfamily\scriptsize, numbers=none]
# start a transformation  
model.addParam("theta") # add a new parameter theta
formode m = model.Mode {
    m.replace(m.flow,"ngps", "2*theta*ngps")
    m.replace(m.flow,"nenc", "2*(1-theta)*nenc")
}
# end of the transformation
\end{lstlisting}
\caption{The third resiliency pattern for the ACC system based on the linear combination of $\nenc$ and $\ngps$.}%
\figlabel{acc_code_3}%
\end{figure}

Alternatively, the \emph{third resiliency pattern}, where we do not need to modify the structure of the original model, is to model the redundancy in the sensory information as a linear combination of different sensor measurements. For example, instead of taking the average of $\ngps$ and $\nenc$, we can model their relationship as $\theta\ngps + (1-\theta)\nenc$, and then synthesize the value of $\theta$ so that the safety property is satisfied. The transformation script of this resiliency pattern is given in~\figref{acc_code_3}. 
%
For this pattern, we assume that a designer still wants to use all sensor measurements even some of them are under spoofing attacks and would like to search for the value of $\theta$ over the range of [0.2, 0.8] (instead of [0, 1]). Given the same attack model for the other patterns, the synthesizer in \toolreaffirm fails to find the value of $\theta$ within the given range to ensure that the safety property is satisfied. However, if we enlarge the range of $\theta$ to [0.1, 0.9], the synthesizer successfully finds the safe value $\theta = 0.1543$. 

\subsection{Single-Machine Infinite-Bus System}
%
Next, we study a class of cyber-physical switching attacks that can destabilize a smart grid system model, and then apply \toolreaffirm to repair the model to provide resilience. A smart power grid system such as the Western Electricity Coordinating Council (WECC) 3-machine, 9-bus system~\cite{sauer1998power}, can be represented as a single-machine infinite-bus (SMIB) system described in \cite{farraj2014practical}. 
%
%
The SMIB system is considered as a \emph{switched system} in which the physical dynamics are changed between two operation modes based on the position of the circuit breaker. The system has two states, $\delta_1$ and $\omega_1$, which are the deviation of the rotor angle and speed of the local generator $G_1$ respectively.
The stability (safety) property of the system can be specified as the following STL formula,
\begin{align} 
	\varphi_{SMIB} & = \Box_{[0, T]} (0 \leq \delta_1[t] \leq 3.5) \wedge (-2 \leq \omega_1[t] \leq 3),\formlabel{smib_stl}
\end{align} 
where $T$ is a simulation duration.  

\vspace{0.5em}
\noindent
{\bf Original SLSF Model.} In this paper, we model the SMIB system as the SLSF model displayed in \figref{smib_plant_model}. The model contains two operation modes whose nonlinear dynamics characterize the transient stability of the local generator $G_1$ presented in~\cite{farraj2014practical}. The transitions between two operation modes depend on the status of the circuit breaker which is connected or disconnected to the load. 
In the model, $\delta_1$ and $\omega_1$ are represented by $delta$ and $omega$, respectively; and the initial conditions are $delta0 \in [0, 1.1198]$ and $omega0 \in [0, 1]$. The discrete variable $load$ captures the open and closed status of the circuit breaker.
\begin{figure}%
	\centering%
    \includegraphics[width=0.8\textwidth]{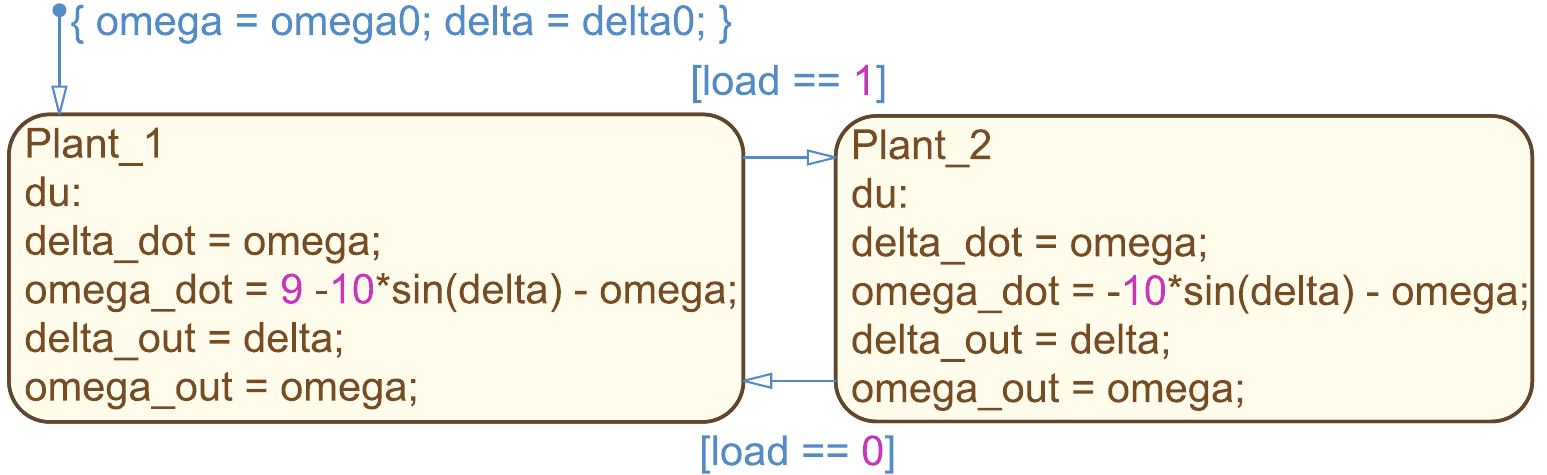}
	\caption{The original SLSF model of the SMIB system.}%
	\figlabel{smib_plant_model}%
\end{figure}%

\noindent
{\bf Sliding-mode attack.} The SMIB system has an interesting property known as a \emph{sliding mode} behavior. This behavior occurs when the state of the system is attracted and subsequently stays within the \emph{sliding surface} defined by a state-dependent switching signal $s(x)\in\Real$~\cite{decarlo1988variable,liu2014coordinated}. When the system is confined on a sliding mode surface, its dynamics exhibit high-frequency oscillations behaviors, so-called a \emph{chattering} phenomenon, which is well-known in the power system design~\cite{sabanovic2004variable}. At this moment, if an attacker forces rapid switching between two operation modes, the system will be steered out of its desirable equilibrium position. As a result, the power system becomes unstable even each individual subsystem is stand-alone stable~\cite{liu2014coordinated}. The sliding-mode attack model of the SMIB system in SLSF format and the unstable behavior of the system under the attack are described in more details in \appref{additional}.  
\begin{figure}[!t]
\begin{lstlisting}[basicstyle=\ttfamily\scriptsize, numbers=none]	
# start a transformation  
model.addParam("theta") # add a new parameter theta
model.addLocalVar("clock") # add a clock variable
formode m = model.Mode {
    m.addFlow("clock_dot = 1")
}
fortran t = model.Trans {
		# a transition only triggers after theta seconds
    t.addGuardLabel("&&","clock > theta") 
		# reset a clock after each transition
    t.addResetLabel("clock = 0") 
}
# end of the transformation
\end{lstlisting}
\caption{A dwell-time resiliency pattern for the SMIB system.}%
\figlabel{smib_code}%
\end{figure}

\begin{figure}[tbp]%
	\centering%
    \includegraphics[width=0.7\textwidth]{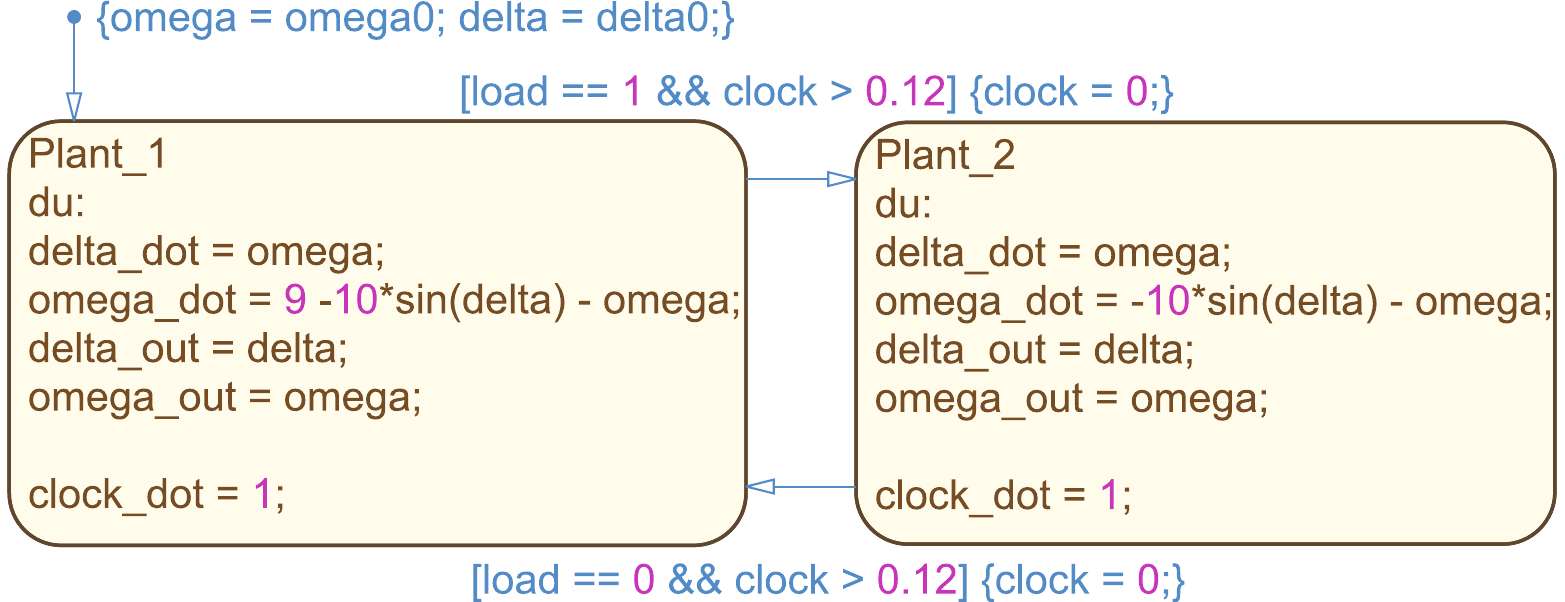}
	\caption{The repaired SMIB model with a synthesized dwell-time.}%
	\figlabel{smib_plant_model_res}%
\end{figure}%

%
%

\vspace{0.5em}
\noindent
{\bf Model Repair for the SMIB system.} 
%
A potential strategy to mitigate a sliding-mode attack is to increase the minimum switching time of the circuit breakers. Indeed, the designer can repair the original model by including a minimum dwell time in each mode of the system to prevent rapid switching.~\figref{smib_code} shows a resiliency pattern written as a HATL script that introduces the \emph{clock} variable as a timer, and the switching time relies on the value of $\theta$.

The model transformation of \toolreaffirm takes the dwell-time pattern shown in~\figref{smib_code}, and then convert the model to a new version that integrates the pattern with the unknown parameter $\theta$.
Then, the model synthesis of \toolreaffirm calls Breach to search for the best (\ie minimum) value of $\theta$ over and the range of $[0, 0.3]$ that ensures the final model satisfies $\varphi_{SMIB}$ (with $T = 10$ seconds) under the sliding-mode attack. The final model, which is stable, is displayed in~\figref{smib_plant_model_res}, where the synthesized value of $\theta$ equals to $0.12$.


\subsection{Missile Guidance System}
\noindent
{\bf Original SLSF Model.} We consider the example of the missile guidance system (MG) provided by Mathworks, which is a good representative of a practical MG system. The original SLSF model has more than 300 blocks. The details of the model can be found at  \url{https://www.mathworks.com/help/simulink/examples/designing-a-guidance-system-in-MATLAB-and-simulink.html}.
In our study, we make a slight modification in the Sensors of the Airframe $\&$ Autopilot subcomponent of the original MG model. In the modified version, we model the missile body rate measurements as an array of two different gyroscopes, and the body rate estimation is obtained by using the average of the measurements obtained from the two gyroscopes. Such modification is reasonable as a practical MG system usually uses an array of gyroscopes to estimate the body rate of a missile. The correctness requirement of the model is that the missile will eventually approach the target where their distance is less than 10.  This requirement can be formulated as an STL formula 
\begin{align} 
	\varphi_{MG} & = \Diamond_{[0, T]} range[t] < 10.\formlabel{mg_stl}
\end{align}   
In the original setting, the MG model satisfies the STL requirement and the noisy levels of two gyroscopes are assumed as $|n_{gyro1}| \leq 0.05$ and $|n_{gyro2} |\leq 0.05$, respectively.

\vspace{0.5em}
\noindent
{\bf Gyroscopes Sensor Attack.} 
The principle of a spoofing attack on the gyroscopes of the MG system is similar to the GPS spoofing attack of the ACC system. In this case, we omit the original assumption $|n_{gyro2}| \leq 0.05$, and employ the new assumption as $|n_{gyro2}| \leq 1$. As a result, the MG model no longer satisfies the STL requirement under this assumption.

\begin{figure}[tbp]
\centering  
\subfigure[Original noise estimator]{\includegraphics[width=0.37\linewidth]{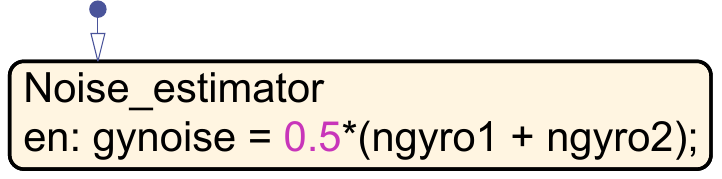}}
\subfigure[Modified estimator using pattern3]{\includegraphics[width=0.53\linewidth]{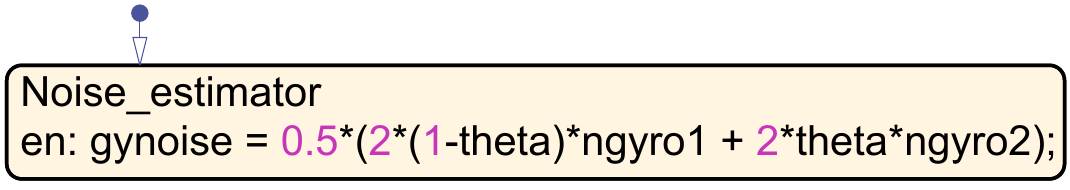}}
\subfigure[Modified estimator using pattern1]{\includegraphics[width=0.34\linewidth]{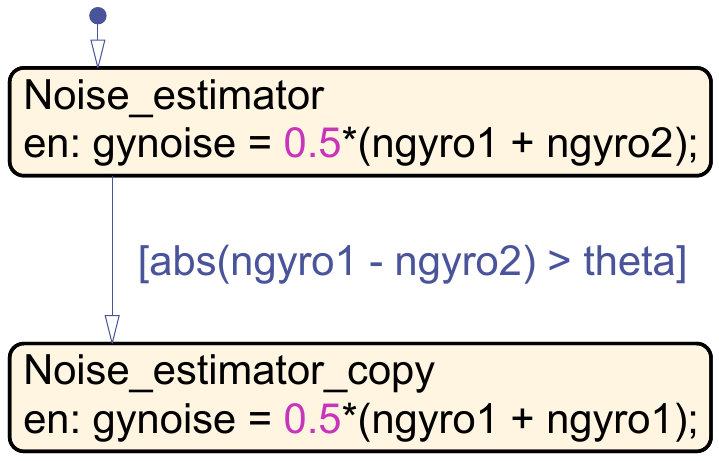}}\hspace{0.2\textwidth}
\subfigure[Modified estimator using pattern2]{\includegraphics[width=0.35\linewidth]{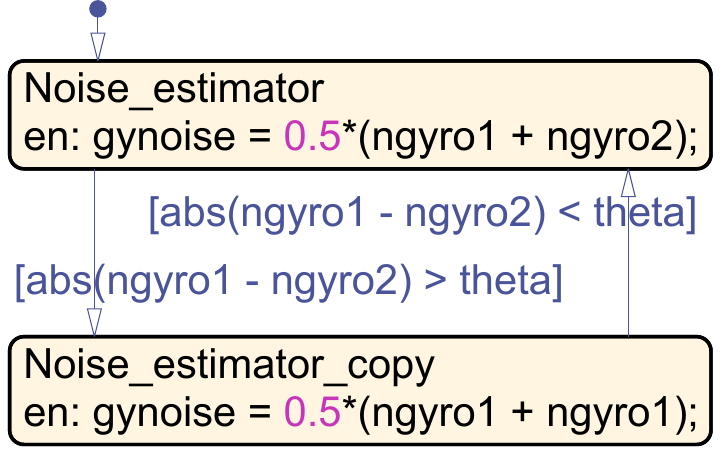}}
\caption{The original and modified noise estimators using three resiliency patterns.}
\figlabel{mg_result}%
\end{figure}

\vspace{0.5em}
\noindent
{\bf Model Repair for the MG System under Gyroscopes Sensor Attack.} To repair the MG model under the gyroscope spoofing attack, we can reuse the three different patterns used to fix the ACC model under GPS spoofing attack. \figref{mg_result} shows the original gyroscope noise estimator which is vulnerable to the spoofing attack and three different repaired versions generated using the three resiliency patterns, respectively. For each resiliency pattern, the synthesized values of $\theta$ and the performance of our Reaffirm toolkit is reported in \tabref{performance_results}.

\section{Related Work}
\seclabel{rw}
%
%
%
%
%

\noindent
{\bf Model-based design of resilient CPSs.} Examples of model-based approaches to ensure resiliency include the approach proposed in~\cite{fitzgerald2012rigorous} that can be used to design a resilient CPS through co-simulation of discrete-event models, a modeling and simulation integration platform for secure and resilient CPS based on attacker-defender games proposed in~\cite{koutsoukos2018sure} with the corresponding testbed introduced in~\cite{neema2018integrated}, the resilience profiling of CPSs presented in~\cite{jackson28resilience}, and the recent works of the design, implementation, and monitor of attack-resilient CPSs introduced in \cite{weimer2018parameter,pajic2017design}.   
Although these approaches can leverage the modeling and testing for a resilient CPS, they do not offer a model repair mechanism or a generic approach to design a resiliency pattern when vulnerabilities are discovered. Our proposed method is complementary to these efforts as we provide a generic, programmable way for a designer to specify a potential edit that can effectively repair the model for improving resiliency.

\vspace{0.5em}
\noindent
{\bf Formal analysis of hybrid systems.} Our approach utilizes Breach to synthesize an SLSF model due to its advantages in performing  falsification, systematic testing and parameter synthesis for hybrid systems. However, Breach cannot give a formal proof of the system correctness. Depending on different types of hybrid systems, other automatic verification tools can be considered to perform a reachability analysis or formally prove whether a system satisfies a given safety property. For examples, d/dt~\cite{asarin2002d} and SpaceEx~\cite{frehse2011cav} are well-known verification tool for linear/affine hybrid systems; Flow*\cite{chen2013flow} and dReach\cite{kong2015dreach} can be used to compute a reachable set of nonlinear hybrid automata; and C2E2 is a verification tool for Stateflow models~\cite{duggirala2015c2e2}. We choose Breach as it is more scalable.
%

\vspace{0.5em}  
\noindent
{\bf Model transformation languages of hybrid systems.} In the context of the model transformation, GREAT is a metamodel-based graph transformation language that can be used to perform different transformations on domain-specific models~\cite{agrawal2003graph,agrawal2003end}. GREAT has been used to translate SLSF models to Hybrid Systems Interchange Format (HSIF)~\cite{agrawal2004semantic}. Such a translation scheme is accomplished by executing a sequence of translation rules described using UML Class Diagram in a specific order. %
Other approaches that also perform a translation from Simulink diagrams to hybrid systems formalisms such as Timed Interval Calculus~\cite{chen2009formal}, Hybrid Communicating Sequential Processes~\cite{liu2010calculus}, Lustre~\cite{tripakis2005translating}, and SpaceEx~\cite{minopoli2016sl2sx}.  
HYST~\cite{bak2015hyst} is a conversion tool for hybrid automata which allows the same model to be analyzed simultaneously in several hybrid systems analysis tools. 
%
However, the problem of designing a scripting language to facilitate transforming models of hybrid systems has not been addressed before.

\section{Conclusion}
\seclabel{conclude}
%
In this paper, we have presented a new methodology and the toolkit \toolreaffirm that effectively assist a designer to repair CPS models under unanticipated attacks automatically. 
The model transformation tool takes a resiliency pattern specified in the transformation language HATL and generates a new model including unknown parameters whose values can be determined by the synthesizer tool such that the safety requirement is satisfied.
We demonstrated the applicability of \toolreaffirm by using the toolkit to efficiently repair CPS models under realistic attacks including the ACC models under the GPS sensor spoofing attack, the SMIB models under the sliding-model attack, and the MG system under gyroscopes spoofing attack.

%
%
\bibliographystyle{splncs04}
\bibliography{luan}
\newpage

\appendix
\section{Appendix: Additional Experimental Results}
\applabel{additional} 
\noindent
{\bf The repaired ACC models.} \figref{acc_model_pat1},  \figref{acc_model_pat2}, and \figref{acc_model_pat3} show the repaired models of the ACC system using the first, second and third resiliency patterns, respectively.
\begin{figure}[!t]%
	\centering%
    \includegraphics[width=0.6\textwidth]{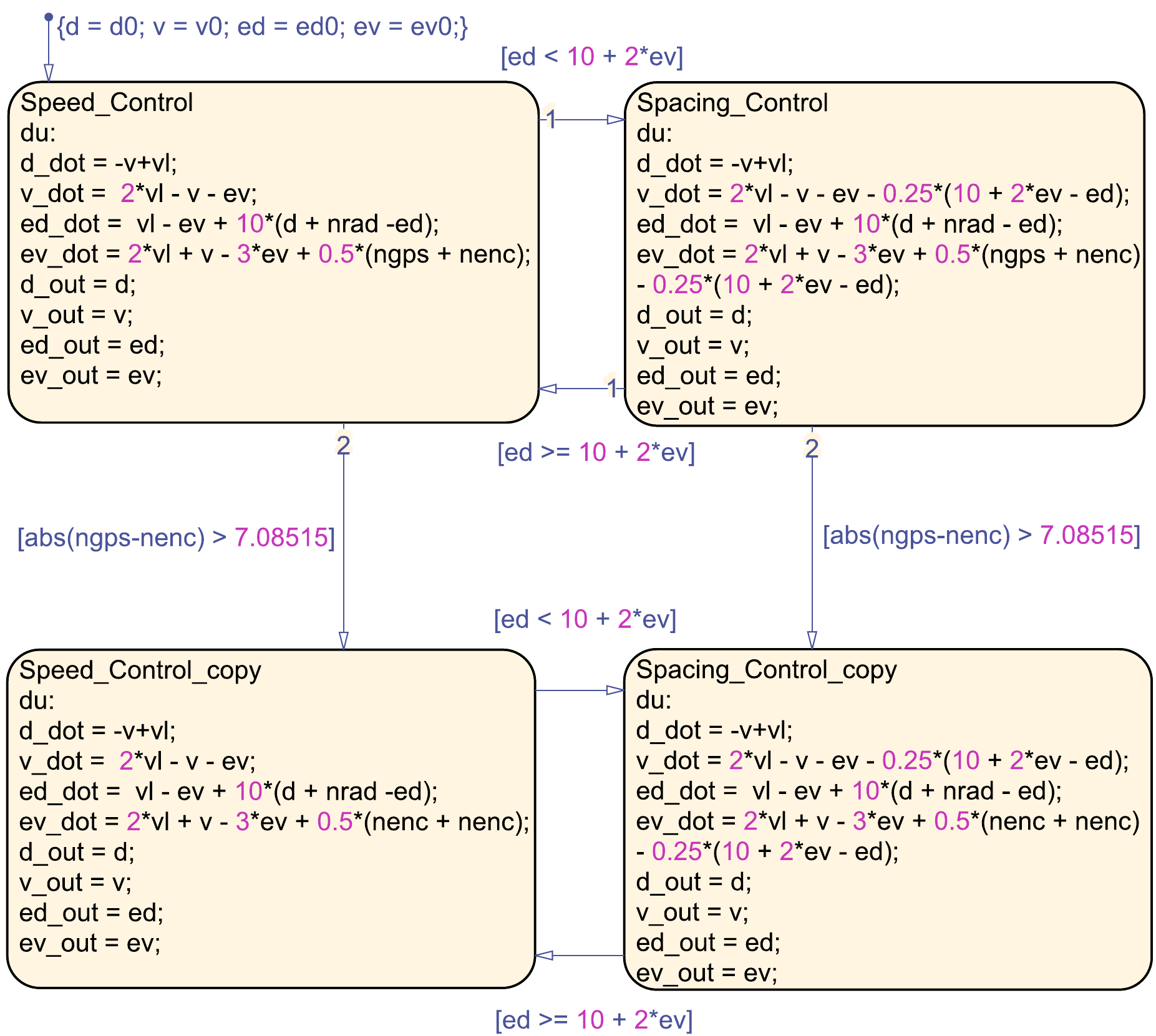}
	\caption{The repaired ACC model with a synthesized value of $\theta = 7.08515$ w.r.t the first resiliency pattern}.%
	\figlabel{acc_model_pat1}%
\end{figure}%
\begin{figure}[!t]%
	\centering%
    \includegraphics[width=0.6\textwidth]{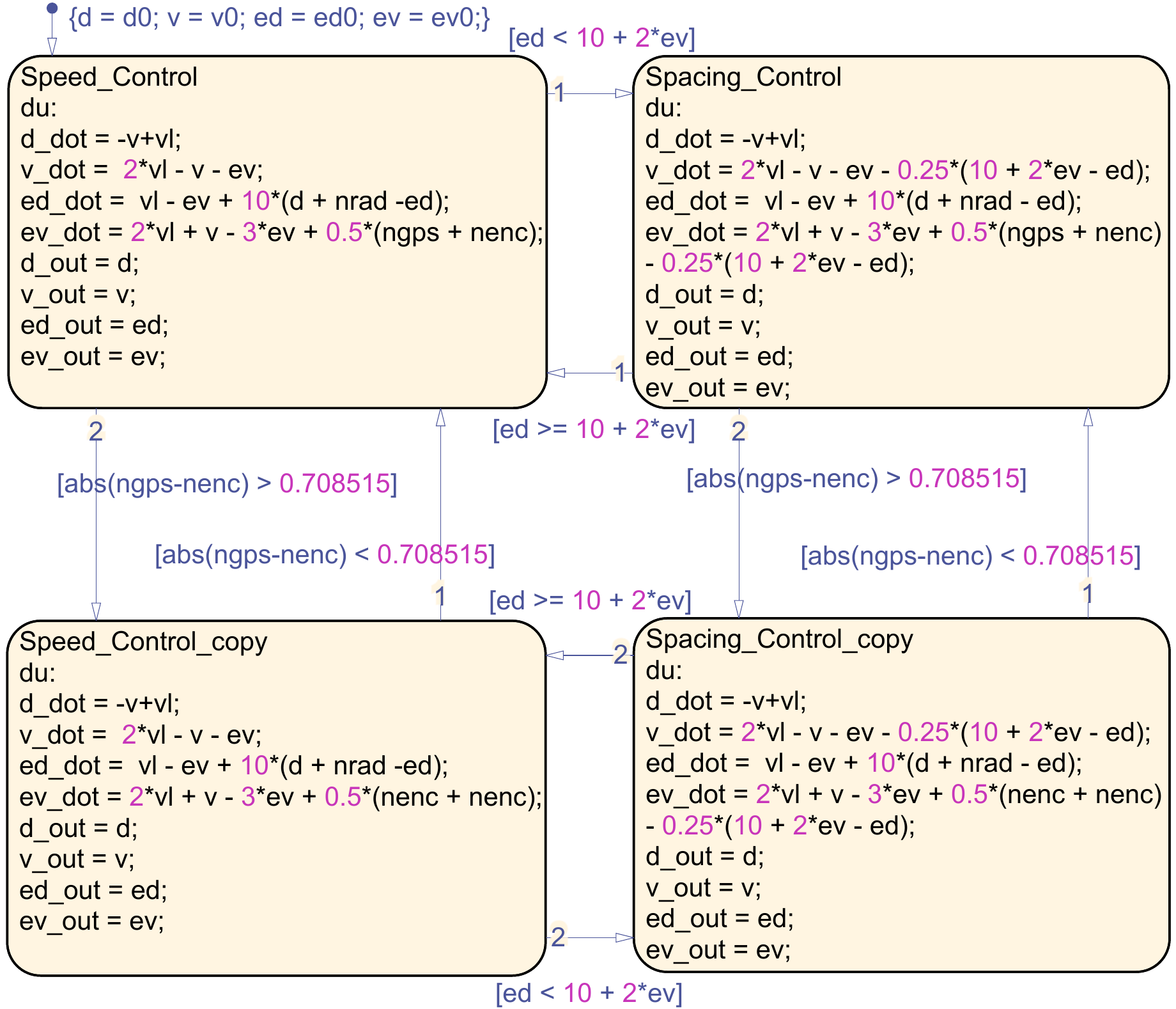}
	\caption{The repaired ACC model with a synthesized value of $\theta = 7.08515$ w.r.t the second resiliency pattern}.%
	\figlabel{acc_model_pat2}%
\end{figure}%
\begin{figure}[!t]%
	\centering
    \includegraphics[width=0.6\textwidth]{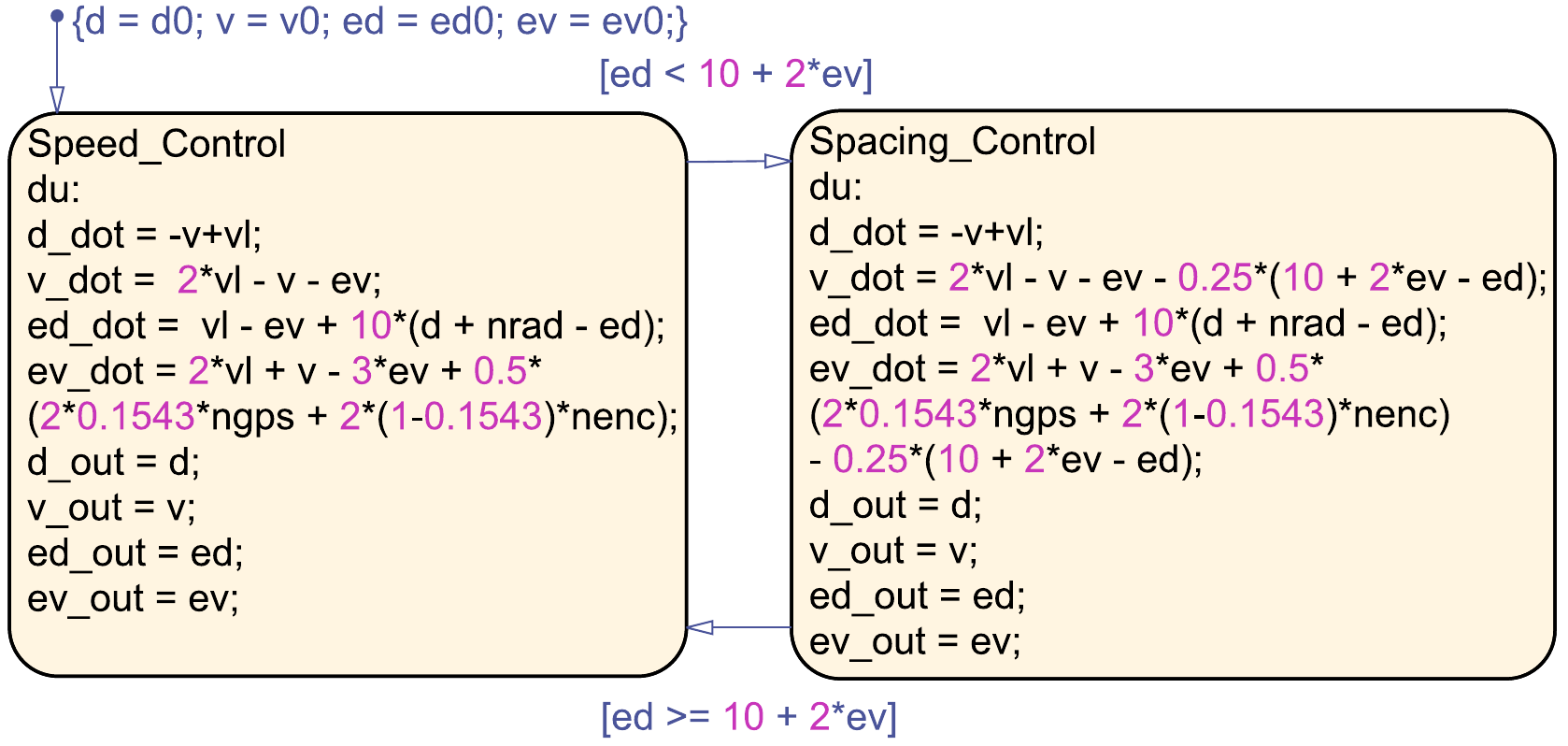}
	\caption{The repaired ACC model with a synthesized value of $\theta = 0.1543$ w.r.t the third resiliency pattern.}%
	\figlabel{acc_model_pat3}%
\end{figure}%

\noindent
{\bf The SMIB system.}
\begin{figure}[t!]%
	\centering%
    \includegraphics[width=0.8\textwidth]{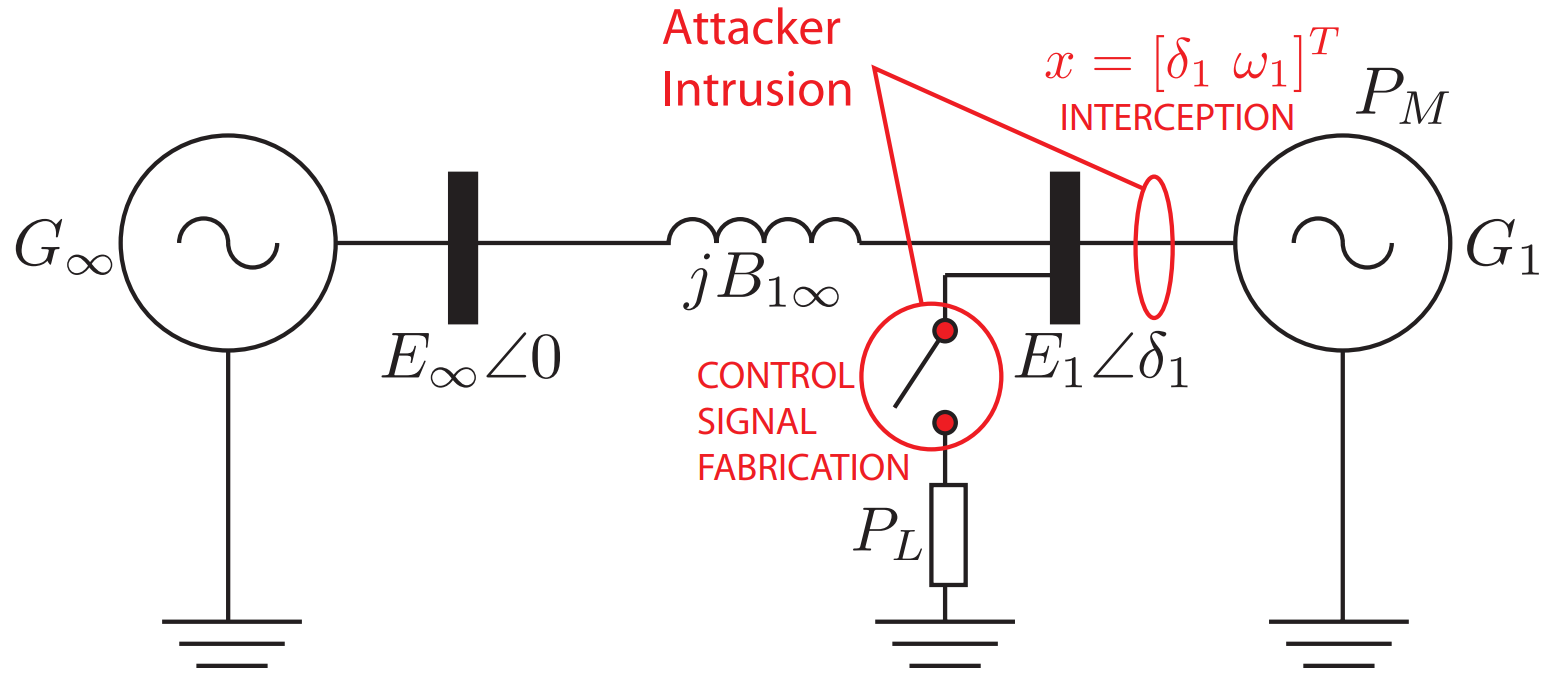}%
	\caption{Single-machine infinite-bus system~\cite{farraj2014practical}.}%
	\figlabel{smib}%
\end{figure}%
%
\figref{smib} represents the general structure of the SMIB system.
In this system, $G_\infty$ and $G_1$ correspondingly represent the SMIB and local generators; $B_\infty$ and $B_1$ denote the infinite and local bus, respectively; $E_\infty$ is the infinite bus voltage; $E_1$ is the internal voltage of $G_1$; $B_{1\infty}$ is the transfer susceptance of the line between $B_1$ and $B_\infty$; and $P_M$ is the mechanical power of $G_1$. The local load $P_L$ is connected or disconnected to the grid by changing a circuit breaker status. 

\noindent
{\bf Sliding-mode attack model of the SMIB system.} To perform the sliding-mode attack on the original SMIB model, we model the attack as the SLSF model shown in \figref{smib_attack_model}. In this model, we assume that the attacker selects a sliding surface $s(x) = \delta_1 + \omega_1 = 0.2$, and the local variable $t$ captures a simulation duration. We note that two transitions from the first mode to the second mode are executed with priorities such that the load is permanently disconnected at some instance where $t \geq 2.5$ seconds. More details of the stages to construct the sliding-mode attack can be found in~\cite{farraj2014practical}. 

\begin{figure}%
	\centering%
    \includegraphics[width=0.8\textwidth]{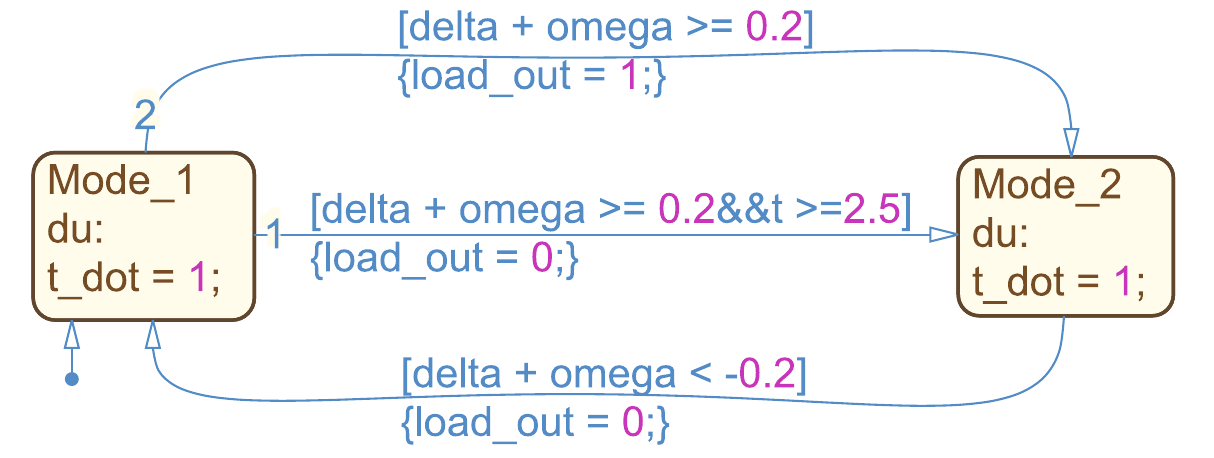}
	\caption{The Stateflow chart models the sliding-mode attack to the SMIB system.}%
	\figlabel{smib_attack_model}%
\end{figure}%

\figref{smib_result} illustrates the examples of stable (\ie without an attack) and unstable (\ie a counterexample appearing under the sliding-model attack) behaviors of the SMIB system returned by running the falsifier of Breach, respectively. The red box defines the stable (safe) operation region of the SMIB system that can be formalized by the STL formula $\varphi_{SMIB}$.

\begin{figure}%
	\centering%
    \includegraphics[width=0.33\textwidth]{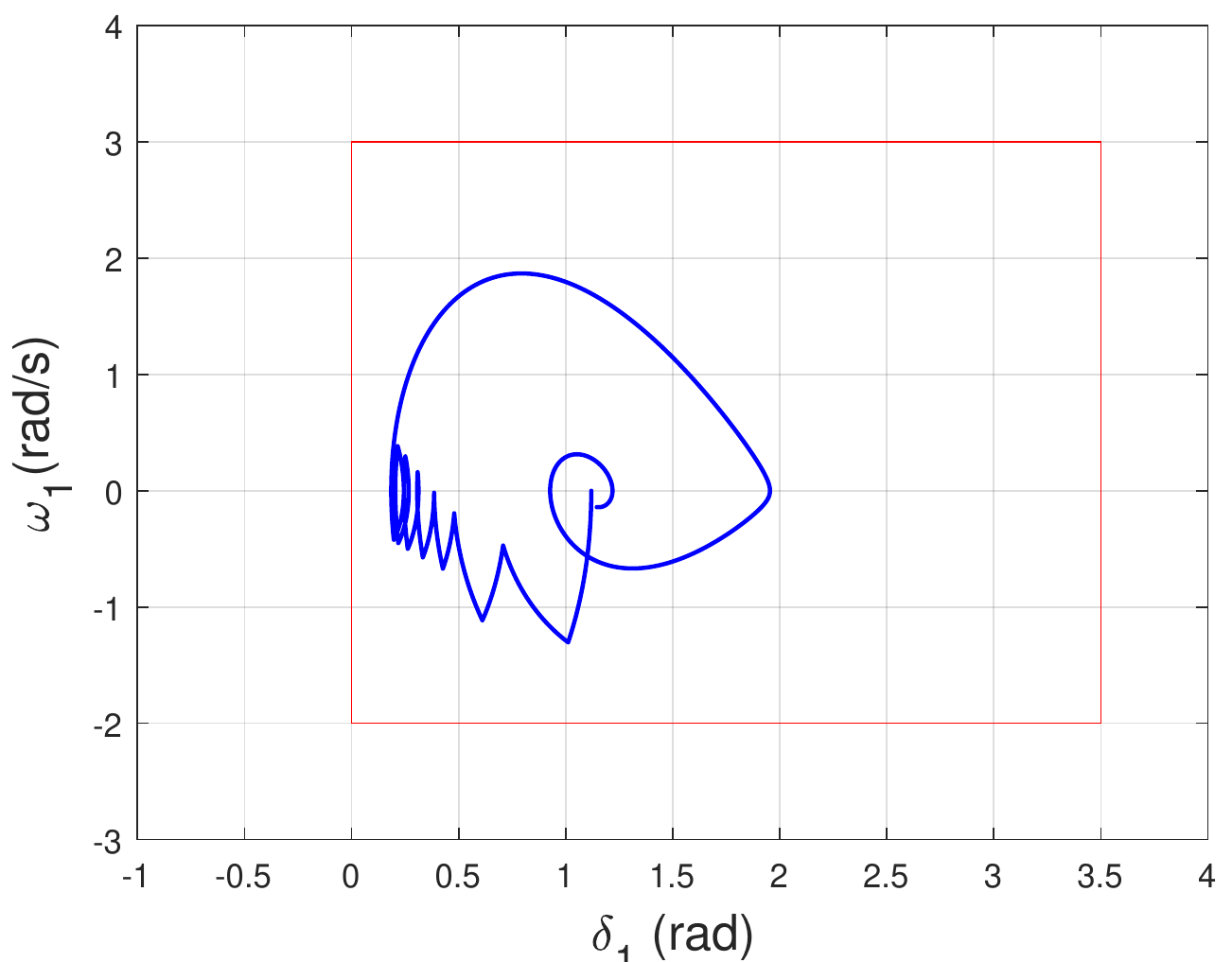}
		\includegraphics[width=0.327\textwidth]{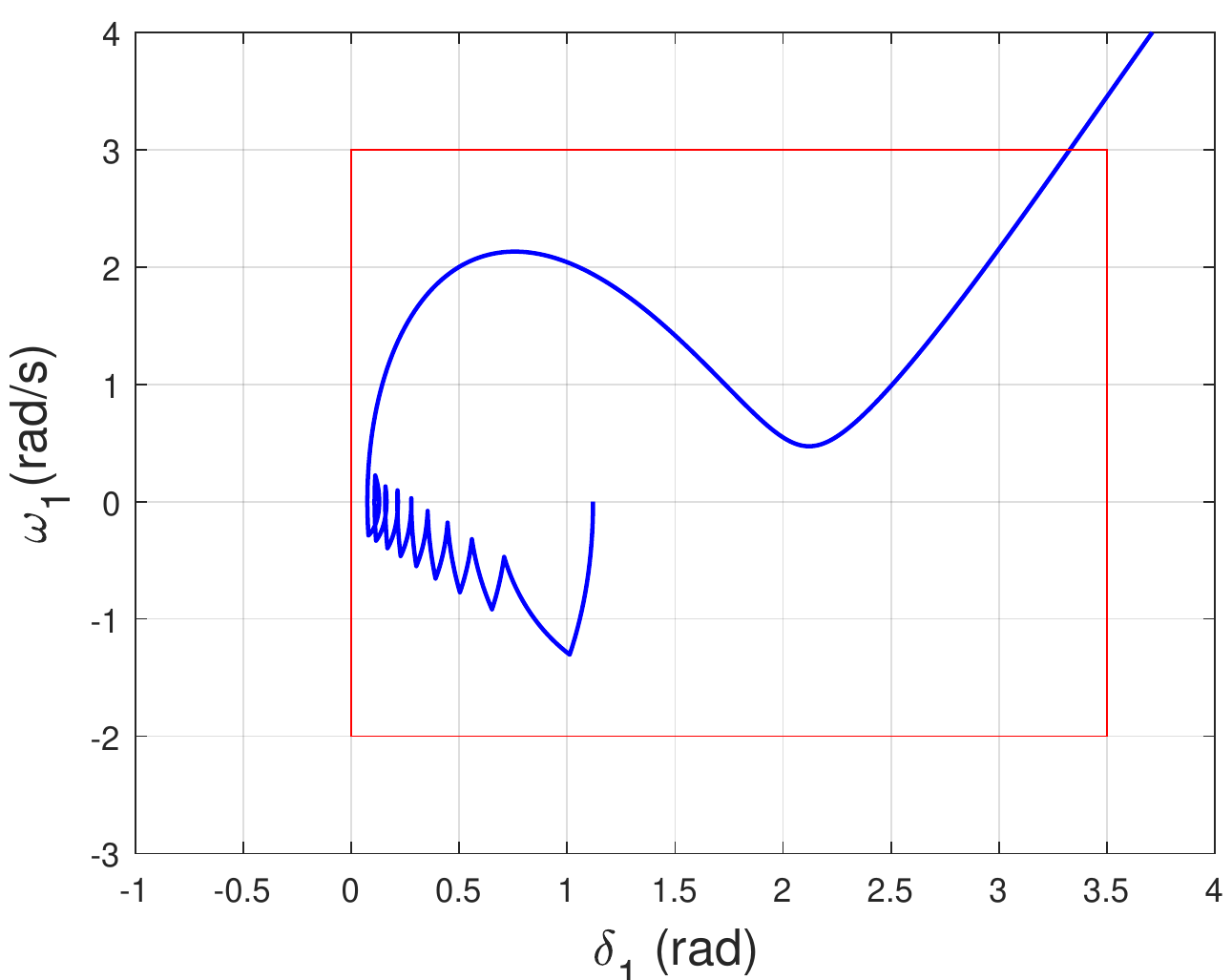}%
		\includegraphics[width=0.314\textwidth]{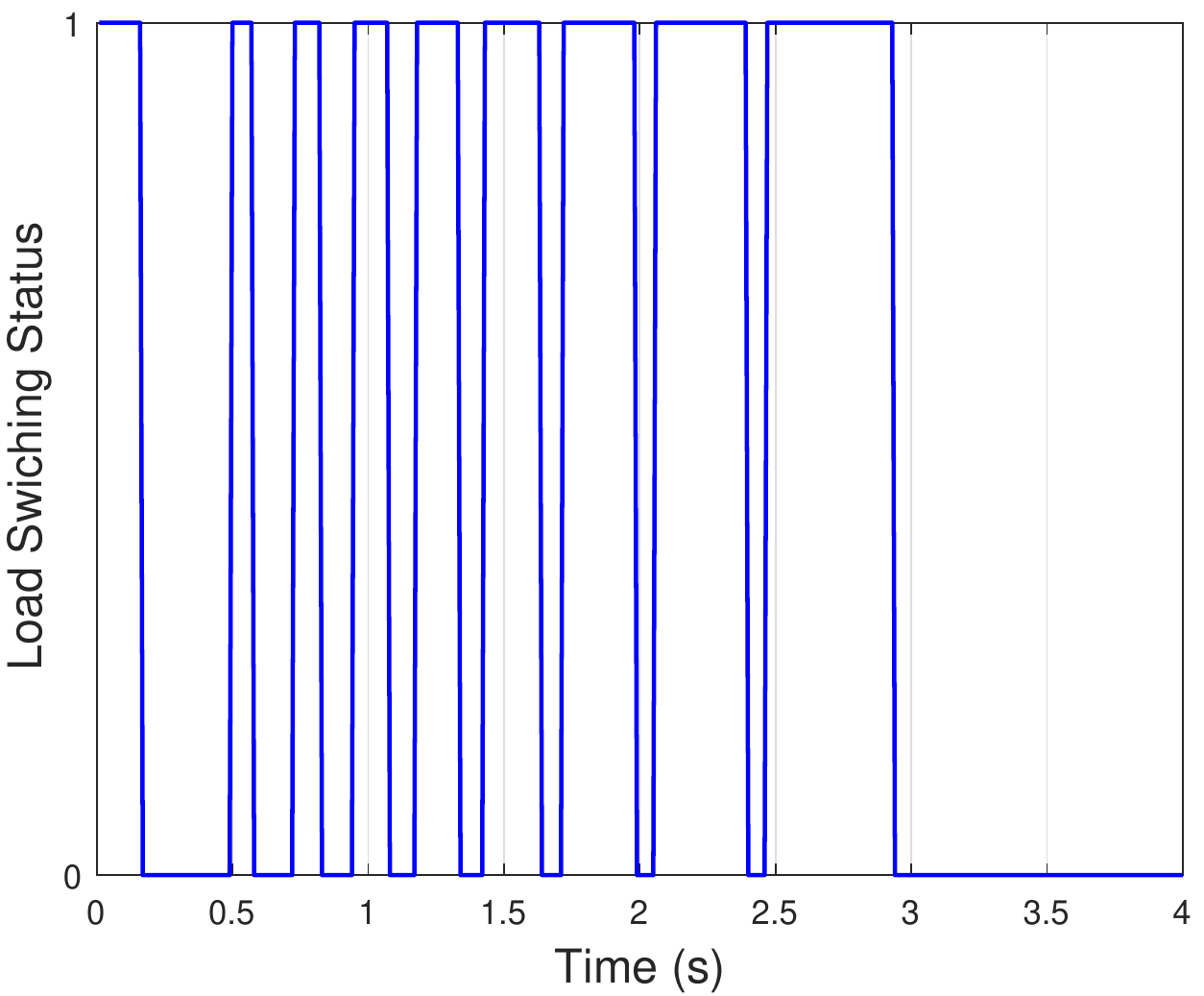}%
	\caption{From left to right: 1) the stable system trajectory without an attack, 2) the counterexample represents the unstable system trajectory under the sliding-mode attack, and 3) the status of a circuit breaker during the attack, where 0 and 1 represent the disconnection and connection of the load $P_L$, respectively.}%
	\figlabel{smib_result}%
\end{figure}%

\end{document}